\documentclass[twocolumn,twocolappendix]{aastex701}

\usepackage{times}
\usepackage[T1]{fontenc}
\usepackage[utf8]{inputenc}
\usepackage{inconsolata}
\usepackage{hyperref}
\usepackage{url}
\usepackage{booktabs}
\usepackage{graphicx}
\usepackage{amsmath}
\usepackage{wrapfig}
\usepackage{lipsum}
\usepackage{tikz}
\usepackage{subcaption}
\usepackage[dvipsnames]{xcolor}
\usepackage{pgfplots}
\usetikzlibrary{positioning, shapes.multipart, backgrounds, fit, math}
\usepackage{soul}
\usepackage{color}
\usepackage[all]{nowidow}
\usepackage{needspace}
\usepackage{float}
\usepackage{lipsum}
\usepackage{listings}
\usepackage{CJKutf8}
\newcommand{\cjk}[1]{\begin{CJK}{UTF8}{gbsn}#1\end{CJK}}

\hypersetup{linkcolor=OliveGreen,citecolor=MidnightBlue,filecolor=cyan,urlcolor=BrickRed}

\newcommand{\teff}{\textrm{T}_{\rm eff}}
\newcommand{\logg}{\textrm{log}\,\textit{g}}
\newcommand{\feh}{\textrm{[Fe/H]}}

\shorttitle{Chemical Cartography Across Time}
\shortauthors{Shen et al.}

\begin{document}

\begin{CJK}{UTF8}{gbsn}

\title{Chemical Cartography Across Time: The Age-Dependent Chemical Structure of the Milky Way with a Homogeneous Sample of 580,000 Giants from Five Surveys}

\author[orcid=0000-0001-6662-7306]{Jeff Shen}
\affiliation{Polymathic AI}
\affiliation{Department of Astrophysical Sciences, Princeton University, 4 Ivy Lane, Princeton, NJ 08540, USA}
\email[show]{shenjeff@princeton.edu}  

\author[0000-0002-6411-8695]{Neige Frankel}
\affiliation{Institute for Astronomy, University of Hawai’i at Mānoa, 2680 Woodlawn Drive, Honolulu, HI 96822, USA}
\email{nfrankel@hawaii.edu}

\author[0000-0003-2866-9403]{David W. Hogg}
\affiliation{Center for Cosmology and Particle Physics, Department of Physics, New York University, 726 Broadway, New York, NY 10012, USA}
\affiliation{Max-Planck-Institut f\"ur Astronomie, K\"onigstuhl 17, D-69117 Heidelberg, Germany}
\affiliation{Center for Computational Astrophysics, Flatiron Institute, 162 Fifth Avenue, New York, NY 10010, USA}
\email{dhogg@flatironinstitute.org}

\author[orcid=0000-0003-2573-9832]{Joshua S. Speagle (\cjk{沈佳士})}
\affiliation{David A. Dunlap Department of Astronomy \& Astrophysics, University of Toronto, 50 St George Street, Toronto, ON M5S 3H4, Canada}
\affiliation{Department of Statistical Sciences, University of Toronto, 9th Floor, Ontario Power Building, 700 University Avenue, Toronto, ON M5G 1Z5, Canada}
\affiliation{Dunlap Institute for Astronomy \& Astrophysics, University of Toronto, 50 St George Street, Toronto, ON M5S 3H4, Canada}
\affiliation{Data Sciences Institute, University of Toronto, 17th Floor, Ontario Power Building, 700 University Avenue, Toronto, ON M5G 1Z5, Canada}
\email{j.speagle@utoronto.ca}

\author[0000-0002-1068-160X]{Shirley Ho}
\affiliation{Polymathic AI}
\affiliation{Center for Computational Astrophysics, Flatiron Institute, 162 Fifth Avenue, New York, NY 10010, USA}
\affiliation{Department of Astrophysical Sciences, Princeton University, 4 Ivy Lane, Princeton, NJ 08540, USA}
\affiliation{Department of Physics \& Center for Data Science, New York University, 726 Broadway, New York, NY 10003, USA}
\email{shirleyho@flatironinstitute.org}

\begin{abstract}

Mapping the distribution of elements in the Milky Way and understanding how that distribution has changed over time can give us insight into the formation and evolution of our Galaxy.
Here we use homogeneously estimated atmospheric parameters, elemental abundances, distances, and ages for ${\sim}580{,}000$ giants from five major spectroscopic surveys---all simultaneously placed on a single, self-consistent scale by one model---allowing us, for the first time, to use a large sample of stars with uniform labels and ages that cumulatively span much of the disk to trace how the metallicity structure of the Milky Way depends on stellar age.
Our chrono-chemical analysis shows that the radial metallicity gradient of the disk depends non-monotonically on age: it is shallow for the youngest stars, steepening to a maximum of ${\sim}{-}0.07$~dex/kpc at intermediate ages (${\sim}5$--$6$~Gyr), and flattening to near zero for the oldest stars. 
This ``U-shaped'' age dependence is driven largely by the low-$\alpha$ (thin-disk) population, with the high-$\alpha$ (thick-disk) sequence having weak gradients at all ages.
In contrast to [Fe/H], the radial [Mg/Fe] gradient is weak ($\lesssim0.01$~dex/kpc) at all ages.
Together, these trends are consistent with an inside-out formation scenario in which the gradient of the star-forming gas has flattened over at least the last ${\sim}6$~Gyr, while the near-zero gradient of the oldest stars reflects their formation in a well-mixed disk, efficient radial migration, or both.

\end{abstract}


\section{Introduction}

The composition of a star is largely determined by the chemistry of the gas cloud that it was born from, and for most elements, this does not change significantly over a star's lifetime \citep{Freeman2002}.
As such, mapping the distribution of stellar abundances across the Galaxy---chemical cartography---along with how this distribution changes with age, traces how the Milky Way has assembled and evolved over cosmic time. 
Such analyses can provide direct constraints on the physical processes that shaped the disk: the inside-out growth of the Galactic disk, radial migration and dynamical heating, and the imprint of past mergers \citep[e.g.,][]{Helmi_2018, Frankel2018, Frankel2019, Belokurov_2022, gaia2023chemicalcartography, Imig_2023}. 
The radial and vertical gradients of metallicity are particularly useful diagnostics, encoding the information about in-situ enrichment, the spatial growth of the disk, and the subsequent redistribution of stars.

Metallicity gradients in the Galactic disk have been measured extensively. 
In the radial direction, the present-day [Fe/H] gradient near the Galactic midplane is negative, of order $-0.06$~dex\,kpc$^{-1}$, and flattens with height above the midplane \citep[e.g.,][]{Cheng2012, Boeche2013dwarf, Hayden2015, Hawkins_2023, Hackshaw2024, Imig_2023}.
In the vertical direction, it is negative and steepens toward the inner Galaxy. 
These trends reflect the superposition of chemically and kinematically distinct populations---the younger, metal-rich, $\alpha$-poor, thin disk and the older, metal-poor, $\alpha$-rich thick disk \citep[e.g.,][]{Gilmore1983}.
Interpretation for age-mixed samples, however, is complicated by the fact that gradients measured for such a sample can correspond to one that is not reproduced by any coeval subsample and can differ systematically from each of them.
The age-mixed radial gradient is shallower, and the vertical gradient steeper, than those of the underlying mono-age populations, due to differences in the spatial distribution of stars of different ages \citep{xiang2015evolutionstellarmetallicitygradients};  
this can obscure the very time evolution that we are interested in.
Where stellar ages are available, the gradients can instead be measured for the underlying mono-age populations forming the sample, revealing that the metallicity gradients themselves evolve with time.

Prior analyses on the evolution of the chemical structure have been limited by two obstacles: lack of reliable ages for large, spatially-extended samples, and the heterogeneity of the surveys that provide the elemental abundances.
Age-resolved measurements of metallicity gradients have thus far come from large single-survey samples or from smaller samples with more precise ages but limited spatial coverage.
\citet{xiang2015evolutionstellarmetallicitygradients} use ${\sim}300{,}000$ LAMOST main-sequence turn-off stars with isochrone ages, but their sample contains no stars younger than ${\sim}2$~Gyr.
\citet{Anders2017} trace the gradient evolution with CoRoT--APOGEE red giants, and \citet{Ciuca_2017} with ${\sim}18{,}000$ TGAS--RAVE dwarfs.
The latter sample is confined to $7.6<R<8.4$~kpc and $|Z|<0.5$~kpc, and so has difficulty separating radial structure from vertical structure.
Expanding the sample by combining surveys has the issue of heterogeneity.
\citet{sun2024mappingchemodynamicsgalacticdisk}, for example, combine LAMOST and APOGEE red clump stars and find that the two pipelines place stars on systematically different scales, so one survey must be recalibrated onto the other post hoc (e.g., by fitting linear corrections) before the two can be combined.
Different spectroscopic surveys probe different regions of the Galaxy with different selection functions and, critically, place abundances on different scales due to differences in processing pipelines, so naively combining the different abundances straight from output catalogs can result in spurious spatial gradients. 
A solution to this will only be more important as the number of surveys increases if we are to exploit the synergies between them \citep[e.g., SDSS-V, MOONS, 4MOST;][]{Kollmeier2017, moonssurvey, 4most}.

Many techniques have been developed to put surveys onto a common scale, each with its own limitation.
\textit{The Cannon} \citep{Ness2015} is a data-driven method that transfers labels from one survey to the labels of another using stars observed in common.
While many have applied this technique to homogenize labels from different surveys, the vast majority do pairwise transfer, and do not make use of more than two surveys at once.
Other works like the \textit{Survey of Surveys} \citep{Tsantaki_2022, Turchi2025} apply post-hoc recalibrations to bring surveys onto a common scale.
\citet{Spina2022} assemble homogenized metallicities from five surveys and measure the radial gradient with ages, but do so for $175$ open clusters, which only sample the disk sparsely.

In this work we address the heterogeneity problem by using the homogeneous catalog of \cite{Shen2026}, in which atmospheric parameters, abundances, distances, and ages for stars from five major spectroscopic surveys are, by design, placed on a single, self-consistent scale by a unified deep learning model \citep{Shen2025} without a need for post-hoc recalibration.
\cite{Shen2026} demonstrate, through extensive internal and external validation, that the resulting catalog has reliable predictions with minimal cross-survey systematics.
We thus are able to---for the first time---use a large sample of stars from multiple major surveys with uniform labels and ages that cumulatively span much of the disk to trace how the metallicity structure of the Milky Way has changed over cosmic time.

The paper is structured as follows. 
In Section~\ref{sec:data} we describe the data sample and our quality cuts.
In Section~\ref{sec:results} we present our analysis of the chemical structure of the Milky Way, and in particular how that structure has evolved over time.
We summarize and discuss our results in Section~\ref{sec:conclusion}. 

\section{Data}
\label{sec:data}

We use the homogeneous catalog of stellar labels from \cite{Shen2026}, and refer the reader to that work for a detailed description of the catalog construction and validation.\footnote{The catalog is publicly available on Zenodo at \protect{\url{https://doi.org/10.5281/zenodo.19830515}}.}
Here we briefly summarize the key aspects relevant to this study.

The catalog is produced by a neural network model based on the architecture of \citet{Shen2025}, which ingests spectra of arbitrary resolution and wavelength range and maps them to a single set of reference labels.
Through extensive validation, \cite{Shen2026} showed that the inferred labels are precise and exhibit minimal cross-survey systematics, yielding a homogeneous catalog of stellar parameters, elemental abundances, ages, and distances derived from five spectroscopic surveys.

We apply quality cuts to retain only stars with well-determined parameters: $\sigma_{\teff}<150$~K, $\sigma_{\logg} < 0.2$~dex, $\sigma_{\feh} < 0.1$~dex, $\sigma_{\log{\rm dist}} < 0.2$~dex, and $\sigma_{\rm age} / {\rm age} < 30\%$.
Since we are primarily interested in stellar ages, we further restrict the sample to the range where age estimates are reliable---specifically, the parameter space covered by APOKASC, which provides the ground-truth age labels for our model. 
This corresponds to $1 < \logg < 3.8$. 
To remove the most egregious outliers, we additionally apply cuts on the out-of-distribution flags provided in the catalog, retaining only stars with \texttt{gmm\_percentile > 0.25}, \texttt{iforest\_percentile > 1}, and \texttt{mahalanobis\_percentile < 99.9}.
Based on the distance calibration analysis in \cite{Shen2026} using open cluster distances, we increase all the distances in the catalog by $0.04~$dex, corresponding roughly to a $10\%$ increase, to correct for a small systematic offset.

We cross-match this sample with the \textit{Gaia} DR3 Source table \citep{GaiaCollaboration2020}, which provides astrometry for ${\sim}$1.8 billion objects, using a 1-arcsecond radius search. 
Prior to the cross-match, we filter the \textit{Gaia} catalog to retain only sources with \texttt{ruwe < 1.4}, to exclude likely unresolved binaries. 
We also restrict the sample to $|Z|<2.5$~kpc. 
Beyond that height, the sample is composed almost entirely of stars from DESI, so its abundance scale cannot be cross-checked against any other; 
further, a majority of those stars have \textit{Gaia} parallaxes that are incompatible with their catalog distances. 
Since our metallicity gradient measurements are made within $|Z|<2$~kpc, this cut does not affect any of our results.

The catalog has one entry per spectrum, so stars observed by more than one survey appear more than once; after all quality cuts we keep one entry per \textit{Gaia} DR3 \texttt{source\_id}, choosing the entry with the smallest [Fe/H] uncertainty and breaking ties by the smallest $\sigma_{\rm age}/{\rm age}$ and then by the fixed survey order APOGEE, GALAH, DESI, SDSS, \textit{Gaia} RVS.
In practice, \cite{Shen2026} has shown that where there is overlap, estimates for the same star from different surveys are consistent with each other (to the level of the measurement uncertainty or better), so the choice of which entry to retain does not affect our results.
This affects 103,872 stars and removes 107,707 entries.
The final sample comprises 581,688 stars.
The number of stars contributed by each survey after all quality cuts and deduplication is listed in Table~\ref{tab:data}.
Throughout this work we use ``SDSS'' to denote the \emph{optical} SDSS spectroscopic surveys---the SDSS Legacy survey, SEGUE, BOSS, and eBOSS---all of which were observed with the optical SDSS/BOSS spectrographs.
APOGEE, a near-infrared survey, is taken with a separate instrument, and we thus treat it as a distinct survey and list it separately in all tables and figures.
This is also the convention followed by The Multimodal Universe \citep{mmu2024} which is the source of the \cite{Shen2026} catalog.

\begin{table}[]
    \centering
    \begin{tabular}{c|ccccc}
    \toprule
        Survey & APOGEE & GALAH & SDSS & DESI & \textit{Gaia} RVS \\
        \midrule
        N & 205,170 & 109,773 & 6,417 & 103,445 & 156,883 \\
     \bottomrule
    \end{tabular}
    \caption{Number of samples that come from each survey, after quality cuts and deduplication. ``SDSS'' here refers to the optical SDSS spectroscopic surveys (Legacy, SEGUE, BOSS, and eBOSS) only.}
    \label{tab:data}
\end{table}

\begin{figure*}[th]
    \centering
    \includegraphics[width=1\linewidth]{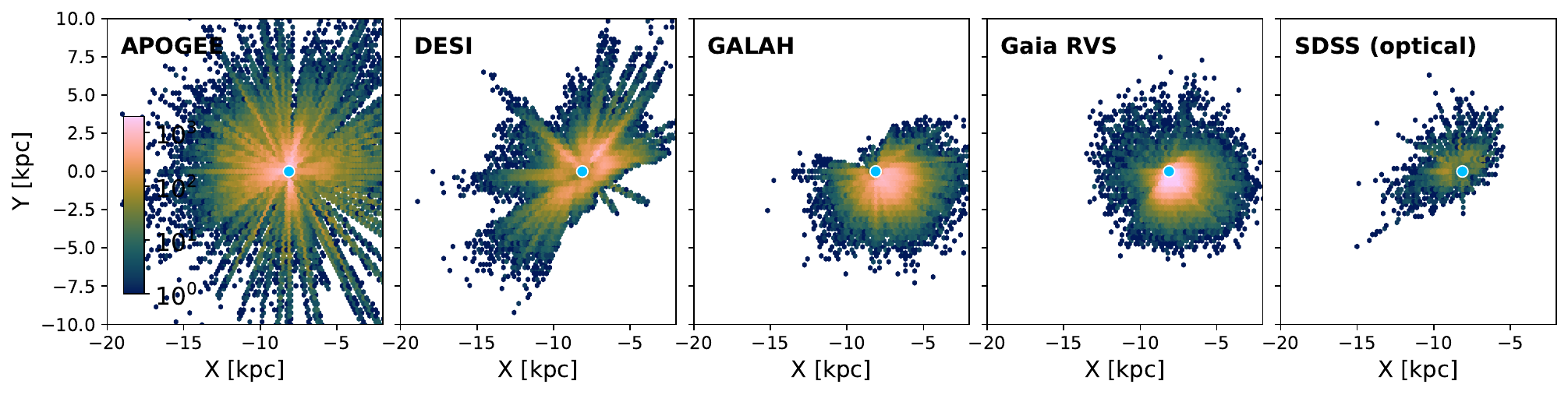}
    \caption{Spatial coverage (X vs Y), coloured by log-sample density, for each survey. The location of the Sun is marked by the blue point.}
    \label{fig:coverage_by_survey}
\end{figure*}

\begin{figure}
    \centering
    \includegraphics[width=1\linewidth]{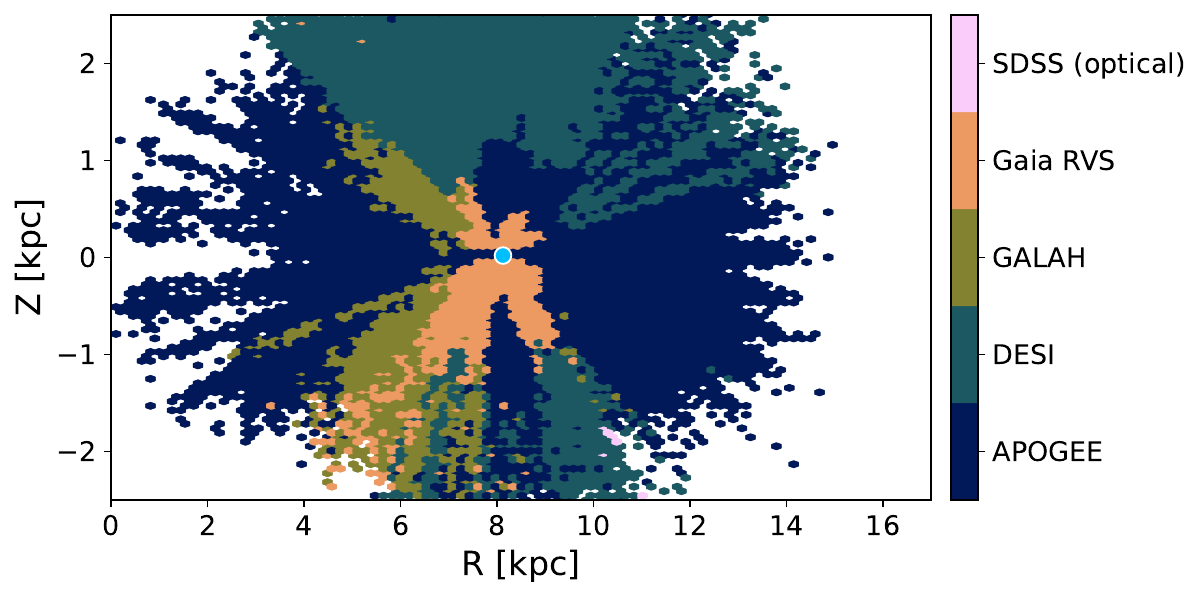}
    \caption{Map of which survey contributes the most stars to each hexbin cell in the $R$--$Z$ plane, for cells with at least five stars.}
    \label{fig:footprint_rz}
\end{figure}

We compute three-dimensional positions using \textit{Gaia} sky coordinates and catalog distances, converting to a Galactocentric reference frame assuming the Sun is located at $R = 8.122$~kpc and $Z = 20.8$~pc above the midplane \citep{astropy:2013, astropy:2018, astropy:2022}.

In Figure \ref{fig:coverage_by_survey} we show a face-on view of the spatial distribution of our selected sample, split by the origin survey. 
This shows the contribution of each of the surveys that have been homogenized to our catalog;
all surveys have high sample counts near the solar neighbourhood, with density rapidly dropping off with distance from the Sun.
Visible in the plot are the pencil-beam patterns of APOGEE and DESI.
Gaia RVS, on the other hand, provides a much more uniform sampling, but only probes nearby stars, whereas DESI is able to probe stars that are ${\sim}10$~kpc or more away from us.
Figure~\ref{fig:footprint_rz} shows which survey contributes the most stars in each region of the $R$--$Z$ plane.

Combining spectroscopic surveys to measure elemental abundance gradients is usually difficult because surveys with systematically different abundance scales target different regions of the Galaxy, as well as different stellar populations; 
the mixture of these heterogeneous surveys can produce spurious abundance gradients in the final data mixture.
We specifically choose the catalog of \cite{Shen2026} to avoid this very problem: because labels for all stars are produced by a single model and placed onto a self-consistent scale, the possibility of differences in the surveys' abundance zero-points masquerading as real structure is eliminated.
However, what remains is the possibility of a spatial sampling bias.
The five surveys that make up our data mixture do not sample the disk uniformly (Figure~\ref{fig:coverage_by_survey}), and each carries its own magnitude limit.
Modelling the combination of five independent survey-specific selection functions is beyond the scope of this paper, and the spatial selection of such magnitude-limited samples is a recognized caveat of survey-based gradient measurements \citep{Imig_2023, gaia2023chemicalcartography}.
The dominant effect is that at large distances, we may preferentially select the most intrinsically luminous (more evolved) giants, which have lower \logg{}. 
We test whether our results are sensitive to our survey selection in Appendix~\ref{apdx:selection} by recomputing the age dependence independently within each survey, and we discuss further limitations in Section~\ref{sec:conclusion}.

\begin{figure}[h]
    \centering
    \includegraphics[width=1\linewidth]{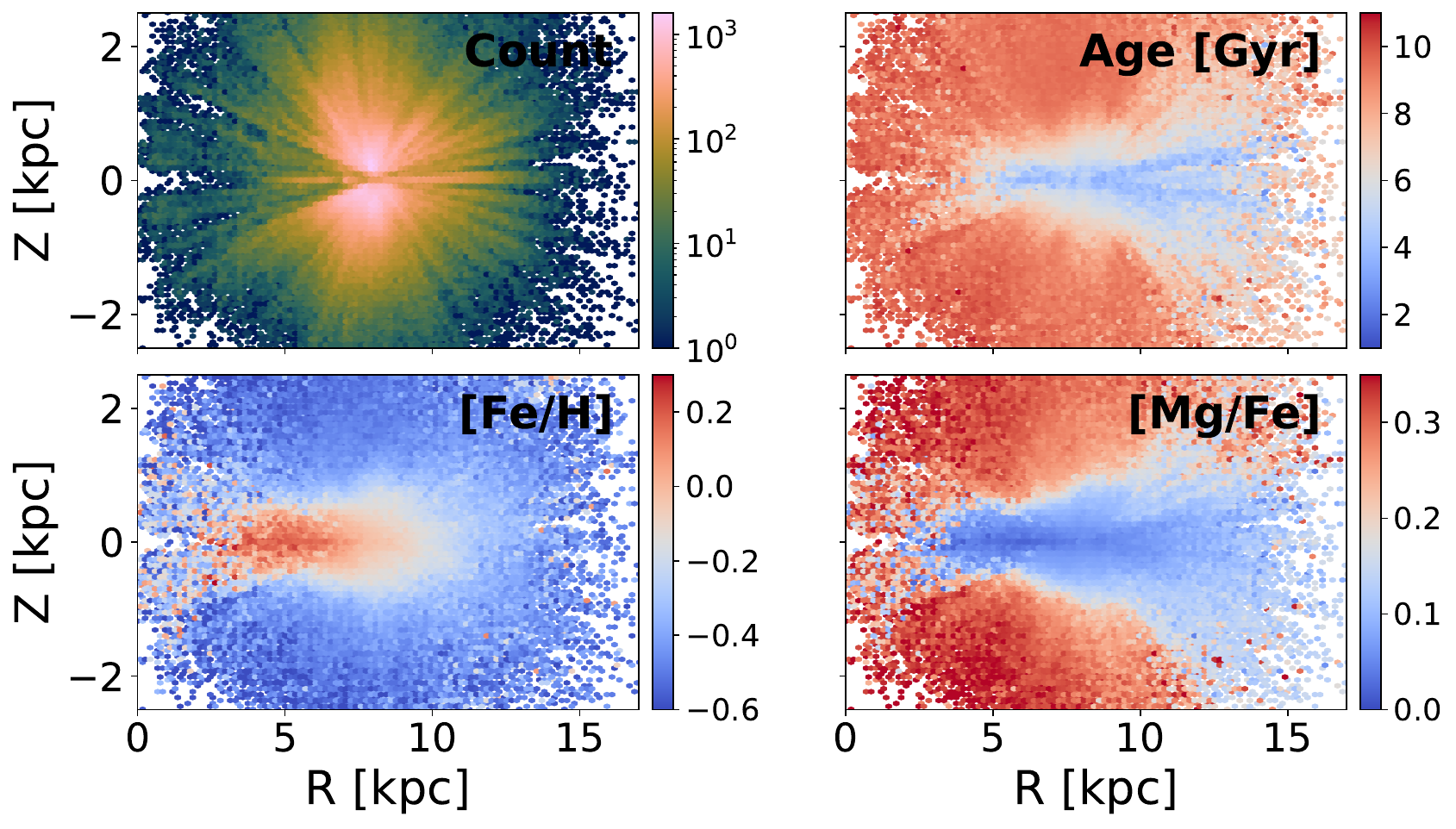}
    \caption{Spatial distribution (R vs Z), coloured by log-sample density (top left), median stellar age in Gyr (top right), median [Fe/H] (bottom left), and median [Mg/Fe] (bottom right).}
    \label{fig:spatial_distribution}
\end{figure}

In Figure \ref{fig:spatial_distribution} we show the spatial distribution of our sample in R vs Z space to better study radial and vertical structure in the Galaxy.
We colour the different panels by sample density (aggregated over all input surveys), age, [Fe/H], and [Mg/Fe].
The structure of the disk is immediately visible: in particular, we see a thin disk of young, metal-rich, $\alpha-$poor stars in the midplane of the galaxy, with older, more metal-poor, $\alpha-$rich stars at larger |Z|.
This transition from a thin layer of young, metal-rich, $\alpha$-poor stars in the midplane into an older, more metal-poor, $\alpha$-enhanced population at larger $|Z|$ is the familiar thin-disk/thick-disk dichotomy \citep{Gilmore1983, Juric2008, Bland-Hawthorn2016}. 
The transition is sharp in the inner Galaxy but becomes progressively blurrier at larger $R$, so that very distinct vertical metallicity and abundance contrast between the two populations washes out in the outer disk.
We quantify this trend for \feh{} in the vertical metallicity gradients below (Figure~\ref{fig:vertical_metallicity_profile}).

Part of this radial change in contrast reflects the flaring of the disk: its vertical scale height increases with Galactocentric radius, so that at large $R$, the young, metal-rich $\alpha$-poor population extends high above the midplane. 
Flaring of the outer disk has been measured from star counts \citep{Momany2006} and, more recently, in mono-age and mono-abundance populations, which are found to flare in both simulations and APOGEE data \citep{Minchev2015, Mackereth2017}. 
Such flaring has been hypothesized to result from processes such as radial migration and satellite infall, and it offers an explanation for the flattening of the vertical metallicity/abundance gradient as we move to larger $R$.

\section{Results}
\label{sec:results}

\subsection{[Fe/H]--[$\alpha$/Fe] plane}

\begin{figure}
    \centering
    \includegraphics[width=\linewidth]{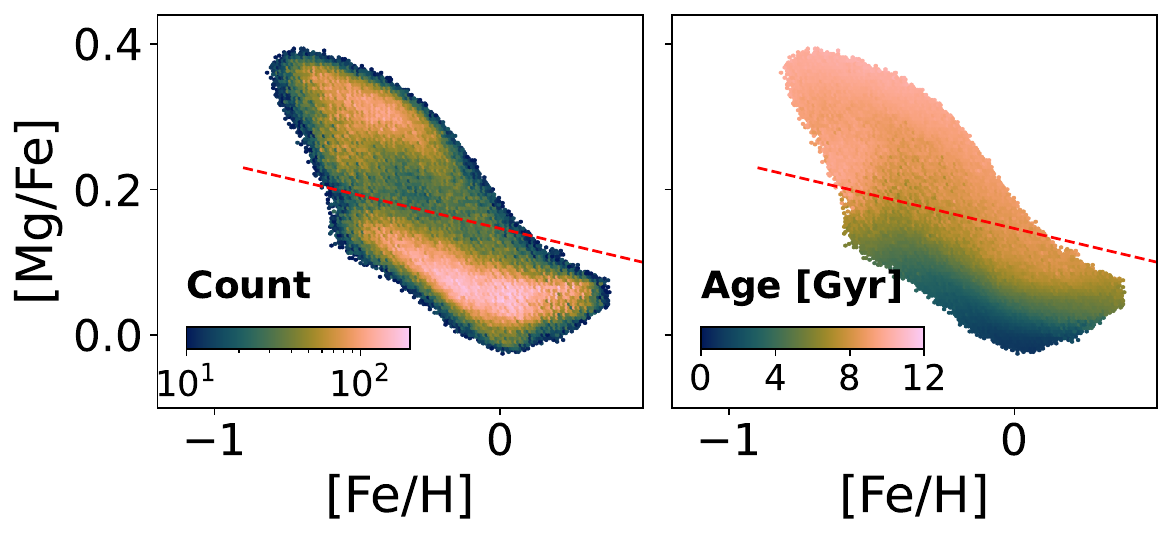}
    \caption{The [Mg/Fe]--[Fe/H] chemical plane. \textbf{Left:} number density on a logarithmic colour scale; the clear bimodality separates the high-$\alpha$ and low-$\alpha$ sequences, and the red dashed line marks the divide we use to classify them. \textbf{Right:} the same plane coloured by median stellar age.}
    \label{fig:mgfe_plane}
\end{figure}

We first show the classic [Mg/Fe] vs [Fe/H] plot in the left panel of Figure \ref{fig:mgfe_plane}, which tracks the star formation history of the Galaxy.
We see a distinct bimodality in the plot, pointing to two distinct formation pathways of the stellar population of the Galaxy. 
In the thick disk, [Mg/Fe] is high and stars are old, suggesting that early in the Galaxy's history, star formation was vigorous, while in the thin disk, Type Ia supernovae have increased [Fe/H] and driven down the [Mg/Fe] ratio over time.
The bimodality and the sense of the two sequences are the classic signature of a delayed iron source: core-collapse (Type~II) supernovae rapidly enrich the interstellar medium in $\alpha$ elements within ${\sim}10^{7}$~yr, whereas Type~Ia supernovae release iron only after a ${\sim}$Gyr delay set by their white-dwarf progenitors \citep{Matteucci1986}.
In the two-infall model \citep{Chiappini1997}, the high-$\alpha$ (thick-disk) sequence forms rapidly from an early gas-accretion episode and the low-$\alpha$ (thin-disk) sequence forms later from a second, slower infall, producing two tracks like those seen here; a revised version calibrated with asteroseismic ages (APOKASC) reproduces the bimodality with a ${\sim}4$~Gyr delay between the two infalls \citep{Spitoni2019}.
An alternate hypothesis attributes the bimodality to a single evolving sequence broadened by radial migration bringing iron-rich stars outward from the inner disk \citep{Schonrich2009}.

In what follows, we will sometimes analyze the thin disk (low-$\alpha$ sequence) and thick disk (high-$\alpha$ sequence) separately.
We thus draw a line through Figure \ref{fig:mgfe_plane} to separate the low-$\alpha$ and high-$\alpha$ sequences.
The entire sample lies in the range $-0.9 \le \mathrm{[Fe/H]} \le 0.5$, and we define the dividing line to be
\begin{equation}
\mathrm{[Mg/Fe]}_{\mathrm{div}} = 0.23 - \frac{0.13}{1.4}\left(\mathrm{[Fe/H]} + 0.9\right),
\end{equation}
which is a straight line of slope ${\approx}\,{-}0.093$.
A star is assigned to the high-$\alpha$ (thick-disk) sequence if $\mathrm{[Mg/Fe]} > \mathrm{[Mg/Fe]}_{\mathrm{div}} + \delta$ (above this line) and to the low-$\alpha$ (thin-disk) sequence if $\mathrm{[Mg/Fe]} < \mathrm{[Mg/Fe]}_{\mathrm{div}} - \delta$ (below this line), with a buffer of $\delta = 0.01$~dex around the line to reduce cross-contamination between the two populations.
This split is purely chemical, with no cut on Galactocentric height $|Z|$. 
Using this separation criterion, we have a sample of 339{,}390 thin disk (low-$\alpha$) stars and 221{,}742 thick disk (high-$\alpha$) stars; 20{,}556 fall inside the buffer zone and are assigned to neither sequence.

The right panel of Figure~\ref{fig:mgfe_plane} shows the same [Mg/Fe]--[Fe/H] plane, but coloured by the median age of the stars in each bin.
We see that the low-$\alpha$ and high-$\alpha$ sequences separate cleanly not only in abundance but also in age. 
The high-$\alpha$ sequence is uniformly old (age $\gtrsim8$--$10$~Gyr), whereas the low-$\alpha$ sequence spans a broad range of ages and includes the youngest, most metal-rich stars (age $\lesssim3$~Gyr).
This is what would be observed if the high-$\alpha$ sequence formed early and rapidly, before Type~Ia supernovae had time to enrich the interstellar medium in iron.
The tightness of the relation between age and location in this plane (and other chemical planes) is precisely what allows for the use of abundance ratios as ``chemical clocks'' \citep[e.g.,][]{Mena2019clock}.

\subsection{Age-resolved metallicity gradients}

We measure metallicity gradients in both the radial and vertical directions, using bins of $R$ from 6 to 15~kpc in 1~kpc increments, $|Z|$ from 0 to 2~kpc in 0.25~kpc increments, and age from 0 to 12~Gyr in 2~Gyr increments.

Keeping in mind the spatial distribution of metallicities and ages (Figure \ref{fig:spatial_distribution}), and that metal-rich, young stars are concentrated at small $|Z|$ while metal-poor, old stars are found at larger $|Z|$: if one does not consider age and pools all samples together, the gradients will be biased.
In the vertical direction, the gradients will be steeper than any individual mono-age population, and in the radial direction, the gradients will be shallower than any individual mono-age population, as also noted by \citet{xiang2015evolutionstellarmetallicitygradients}.

Furthermore, gradients depend on both $R$ and $|Z|$, making it important to consider both directions simultaneously.
When examining vertical gradients, one should work at fixed $R$, and vice versa; otherwise the effects are entangled and trends detected in, for example, $|Z|$ may in fact be driven by radial effects.
This warning was also raised by \citet{Cheng2012}.

When we trace the age dependence of the gradients below, we use finer age bins of $1$~Gyr, rather than the $2$~Gyr bins used elsewhere, to better resolve the temporal trends, and for simplicity (keeping in mind the warning above) we analyze only stars with $|Z|<1$~kpc and report a single radial gradient per age, fit to $1$~kpc radial bins over $6<R<14$~kpc.

\begin{figure*}[th]
    \centering
    \includegraphics[width=0.85\linewidth]{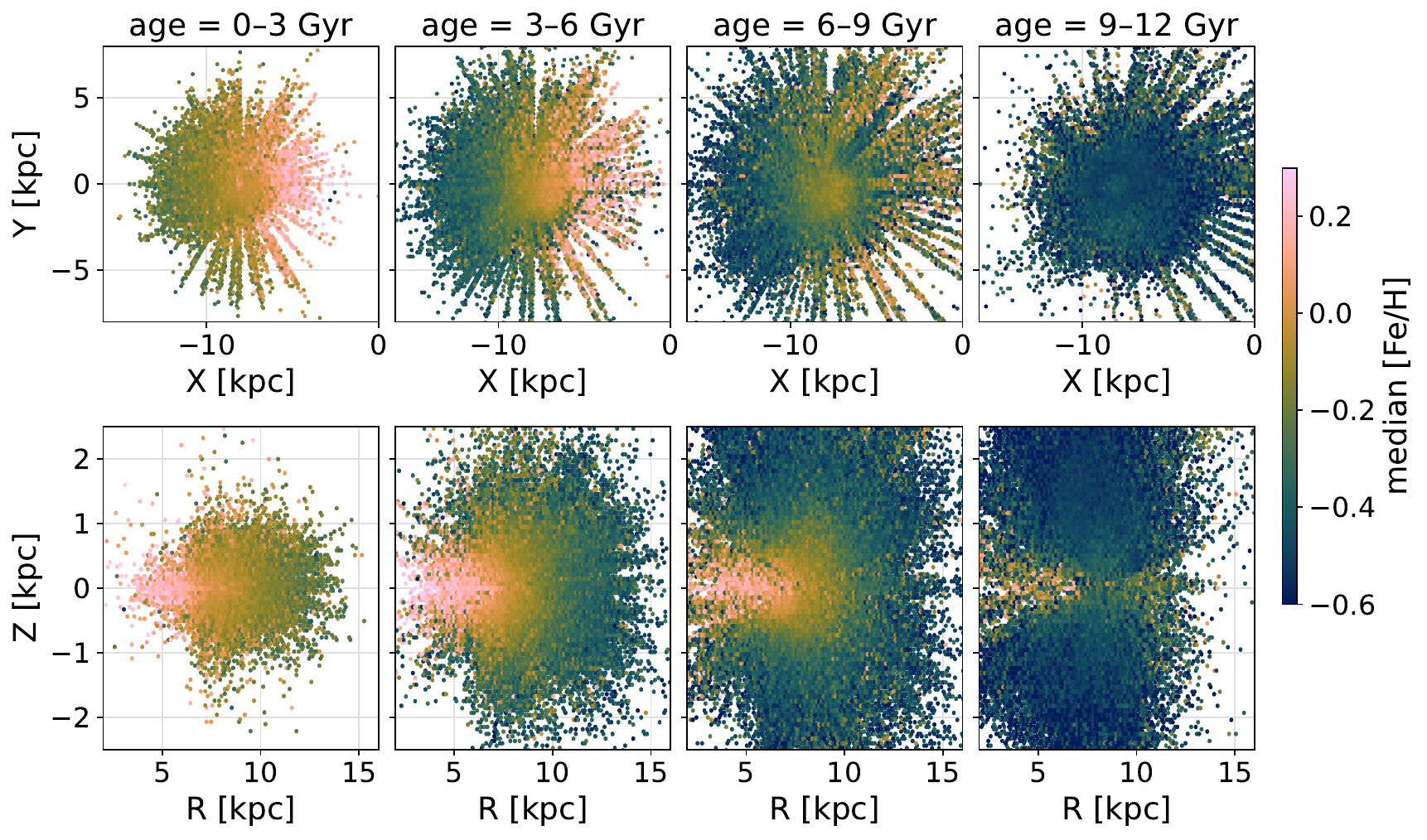}
    \caption{Median [Fe/H] in the face-on $X$--$Y$ plane (top row) and edge-on $R$--$Z$ plane (bottom row), in four bins of stellar age (columns, youngest on the left). The Sun is at $X\approx-8.1$~kpc.}
    \label{fig:feh_maps_by_age}
\end{figure*}

Before we quantify anything, we show in Figure~\ref{fig:feh_maps_by_age} an overview of how the metallicity distribution has evolved over time.
We show the median [Fe/H] within a small spatial hexbin in both face-on ($X$--$Y$) and edge-on ($R$--$Z$) views across four bins in age. 
We show every cell that contains at least one star.
A few trends stand out.
First, in general, at fixed age the disk is more metal-rich toward the Galactic center and more metal-poor in the outskirts, i.e., there is a negative radial gradient. 
Second, this gradient is present in the three youngest age bins, with the largest inner-to-outer contrast at intermediate ages, but washes out for the oldest stars: by age $>9$~Gyr the sample is nearly uniformly metal-poor at $\feh\approx-0.5$ across the entire radial range. 
The oldest populations are also markedly more vertically extended, reflecting the transition from a young, thin, radially-structured disk to an old, thick, well-mixed component \citep[e.g.,][]{Imig_2023}.

\subsection{Age-resolved radial metallicity gradients}

\begin{figure*}[th]
    \centering
    \includegraphics[width=1\linewidth]{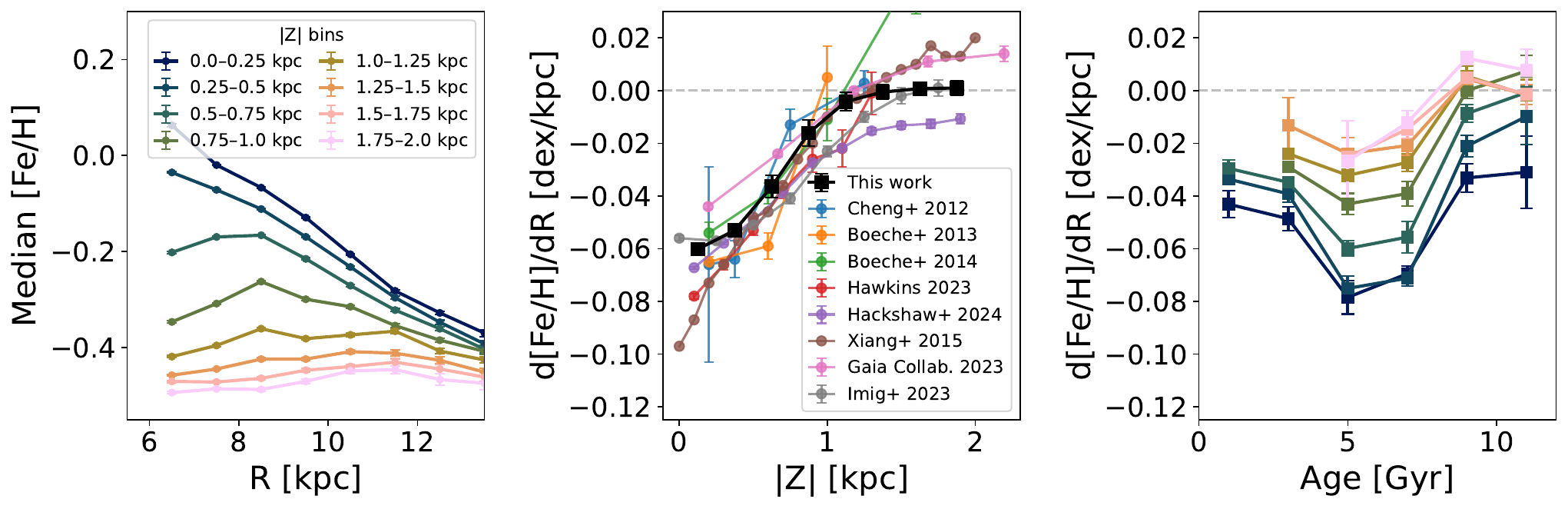}
    \caption{Radial metallicity profile and gradient. \textbf{Left}: median [Fe/H] versus $R$ in bins of $|Z|$. \textbf{Middle}: the radial gradient as a function of $|Z|$ (black), compared to literature measurements. \textbf{Right}: the radial gradient as a function of age, in bins of $|Z|$.}
    \label{fig:radial_metallicity_profile}
\end{figure*}

We show the radial metallicity structure of the Galaxy in Figure \ref{fig:radial_metallicity_profile}.
The left panel shows the median [Fe/H] as a function of $R$ in bins of $|Z|$.
Close to the midplane (small $|Z|$), the median metallicity at small $R$ starts off high but rapidly drops off as we move away from the Galactic center.
At larger $|Z|$, the metallicity profile is relatively flat at $\feh\approx-0.45$.

These results agree very well with what is observed by \citet{Imig_2023}, who use ${\sim}66{,}500$ red giants from APOGEE to map the chemical structure of the Galaxy.
The authors observe that at small Galactocentric radii ($R\approx5~$kpc), the median [Fe/H] measurement at $|Z|=0~$kpc is ${\sim}0.15~$dex, whereas at $|Z|=1.7~$kpc it is ${\sim}-0.5~$dex.
Moving out to larger radii, all $|Z|$ bins converge to $\feh\approx-0.4~$dex at $R=13~$kpc, similar to what we see in Figure \ref{fig:radial_metallicity_profile}.
They also observe that for intermediate $|Z|$ bins, there appears to be a peak in metallicity around $R\approx8~$kpc.
 
The middle panel shows the radial metallicity gradient, as determined by linear fits to the trends in the left panel, as a function of $|Z|$.
All measurements require more than 50 samples per bin and at least 3 valid bins for the linear fit.
Our result across the entire sample is shown in black;
we see a steep decrease in the strength of the gradient (i.e., increase in the gradient) out to $|Z|\approx1.3~$kpc, beyond which there is a flat or very shallow increase in the (signed) gradient.
We measure a negative radial metallicity gradient at small $|Z|$ of ${\rm d[Fe/H]/dR}=-0.060\pm0.002$, flattening to values consistent with zero beyond $|Z|\approx1.3$~kpc (${\rm d[Fe/H]/dR}=0.001\pm0.002$ in the $1.50$--$1.75$~kpc bin).

We compare our measurements to those from the literature \citep{Cheng2012, Boeche2013dwarf, Boeche2014giant, Hawkins_2023, Hackshaw2024, xiang2015evolutionstellarmetallicitygradients, gaia2023chemicalcartography, Imig_2023} and find, in general, similar trends.
Although the literature values disagree with one another by more than their quoted uncertainties, this is due to systematic errors not being accounted for in the uncertainties.
For example, the selection of the stellar population being used to derive the gradients can have an effect on the measured value of the gradient.
In the literature, \citet{Hawkins_2023} find a negative radial metallicity gradient of ${\rm d[Fe/H]/dR}=-0.078\pm0.001$ at $|Z|=0.1~$kpc, flattening out to ${\rm d[Fe/H]/dR}=-0.001\pm0.008$ (consistent with zero, or even a slightly positive gradient) at $|Z|=1.3~$kpc in a linear fashion with respect to $|Z|$.
\citet{Cheng2012} use $7010$ MSTO stars from SEGUE to measure radial metallicity gradients, finding that ${\rm d[Fe/H]/dR}=-0.066_{-0.044}^{+0.030}$ for stars with $0.15~{\rm kpc}<|Z|<0.25~{\rm kpc}$ and increasing to a slight positive gradient of ${\rm d[Fe/H]/dR}=0.0028_{-0.0052}^{+0.0071}$ for stars with $1.0~{\rm kpc}<|Z|<1.5~{\rm kpc}$.
\citet{Boeche2013dwarf}, using $20{,}000$ dwarfs from RAVE, find that ${\rm d[Fe/H]/dR}=-0.065\pm0.002$ for stars at $|Z|<0.4~$kpc, flattening out to ${\rm d[Fe/H]/dR}=0.005\pm0.012$ for stars at $|Z|>0.8~$kpc.
In a followup work, \citet{Boeche2014giant} use $18{,}000$ red-clump stars from RAVE and find similar trends: a negative gradient of $-0.054 \pm 0.004$ for $|Z|<0.4~$kpc, increasing to a positive $0.047\pm0.018$ for $|Z|>1.2~$kpc.
We also compare to the results of \citet{xiang2015evolutionstellarmetallicitygradients}, for which we make a best-effort estimate of the measurements since no exact numbers are reported;
here the metallicity gradient is strongly negative at small $|Z|$, at ${\rm d[Fe/H]/dR}\approx-0.097$, increasing roughly linearly out to a flat gradient at $|Z|\approx1.3~$kpc before increasing more slowly with $|Z|$.
This general trend is also consistent with the results from \citet{gaia2023chemicalcartography} for their giant sample: a negative radial gradient at small $|Z|$, increasing to a slight positive radial gradient at large $|Z|$.
Our exact numbers are in Table \ref{tab:radial_gradients}.
In general, our results are in very good agreement with the literature measurements across the full range of |Z| for which we have measurements.

\begin{table}
  \centering
  \caption{Radial metallicity gradients as a function of $|Z|$.}
  \label{tab:radial_gradients}
  \begin{tabular}{cccc}
    \hline
    $|Z|_{\rm min}$ [kpc] & $|Z|_{\rm max}$ [kpc] & $N$ & d[Fe/H]/d$R$ [dex/kpc] \\
    \hline

    0.00 & 0.25 & 109373 & $-0.0602 \pm 0.0023$ \\
    0.25 & 0.50 & 150990 & $-0.0530 \pm 0.0013$ \\
    0.50 & 0.75 & 109539 & $-0.0364 \pm 0.0043$ \\
    0.75 & 1.00 & 71742 & $-0.0161 \pm 0.0049$ \\
    1.00 & 1.25 & 48494 & $-0.0042 \pm 0.0035$ \\
    1.25 & 1.50 & 33967 & $-0.0005 \pm 0.0027$ \\
    1.50 & 1.75 & 23237 & $0.0008 \pm 0.0023$ \\
    1.75 & 2.00 & 15944 & $0.0010 \pm 0.0027$ \\
    \hline
  \end{tabular}
\end{table}

\begin{figure*}[th]
    \centering
    \includegraphics[width=1\linewidth]{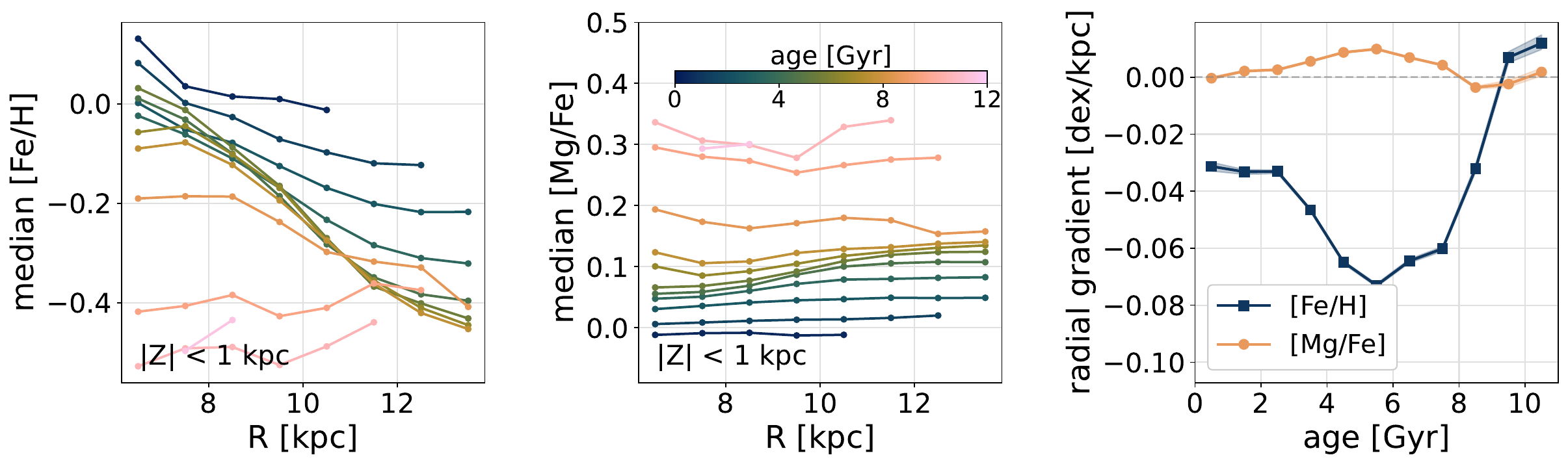}
    \caption{Evolution of the radial abundance gradients. \textbf{Left:} median [Fe/H] versus $R$ for mono-age populations near the midplane ($|Z|<1$~kpc), coloured by age. \textbf{center:} the same for [Mg/Fe], whose profiles are nearly flat and simply stacked by age. \textbf{Right:} the fitted radial gradient as a function of age for [Fe/H] (dark blue squares) and [Mg/Fe] (orange circles); the shaded bands show the $16$--$84$th percentile Monte Carlo scatter about the fiducial curve, from resampling the per-star distance, age, and abundance uncertainties ($N=200$ realizations, independent Gaussians, distance in log-space). The [Fe/H] gradient is shallow for the youngest stars, steepens to a maximum at age $\approx5$--$6$~Gyr, and flattens to near zero for the oldest stars, whereas the [Mg/Fe] gradient is weak at every epoch.}
    \label{fig:radial_gradient_vs_age}
\end{figure*}

The radial gradient is itself a function of height.
The right panel of Figure~\ref{fig:radial_metallicity_profile} shows that ${\rm d[Fe/H]/d}R$ is steepest and most strongly age-dependent near the midplane and grows progressively shallower and less age-dependent at larger $|Z|$, until in the thick disk ($|Z|\gtrsim1$~kpc) it is weak and essentially flat at every age; the middle panel of the same figure shows the corresponding monotonic weakening of the gradient with $|Z|$.
To show the age dependence of the radial gradient near the plane in a different way, we present in Figure \ref{fig:radial_gradient_vs_age} a midplane-averaged view which considers all stars within $|Z|<1$~kpc and measures the radial gradient in finer $1$~Gyr age bins. 
The left panel shows the mono-age radial [Fe/H] profiles, coloured by age, and the rightmost panel shows the fitted [Fe/H] gradient as a function of age (blue curve; the middle panel and the [Mg/Fe] curve in the right panel are discussed in Section~\ref{sec:mgfe_gradient} below).
The shaded band in the rightmost panel is the $16$--$84$th percentile from a Monte Carlo resampling of the per-star distance, age, and [Fe/H] uncertainties ($N=200$ realizations, each perturbing the three quantities by independent Gaussians); it is far narrower than the amplitude of the trend, and the median of the realizations is within $0.002$~dex\,kpc$^{-1}$ of the fiducial curve at every age.
This test therefore shows that the observed trend is stable to added noise at the level of the catalog uncertainties (the median age uncertainty is $0.2$--$0.4$~Gyr); We examine larger/systematic age errors with an injection test in Appendix~\ref{apdx:ages}.
We further verify that the measured age dependence is not an artifact of survey selection by recomputing it independently for each of the five constituent surveys and showing that all recover the same shape (Appendix~\ref{apdx:selection}).

\begin{figure}
    \centering
    \includegraphics[width=1\linewidth]{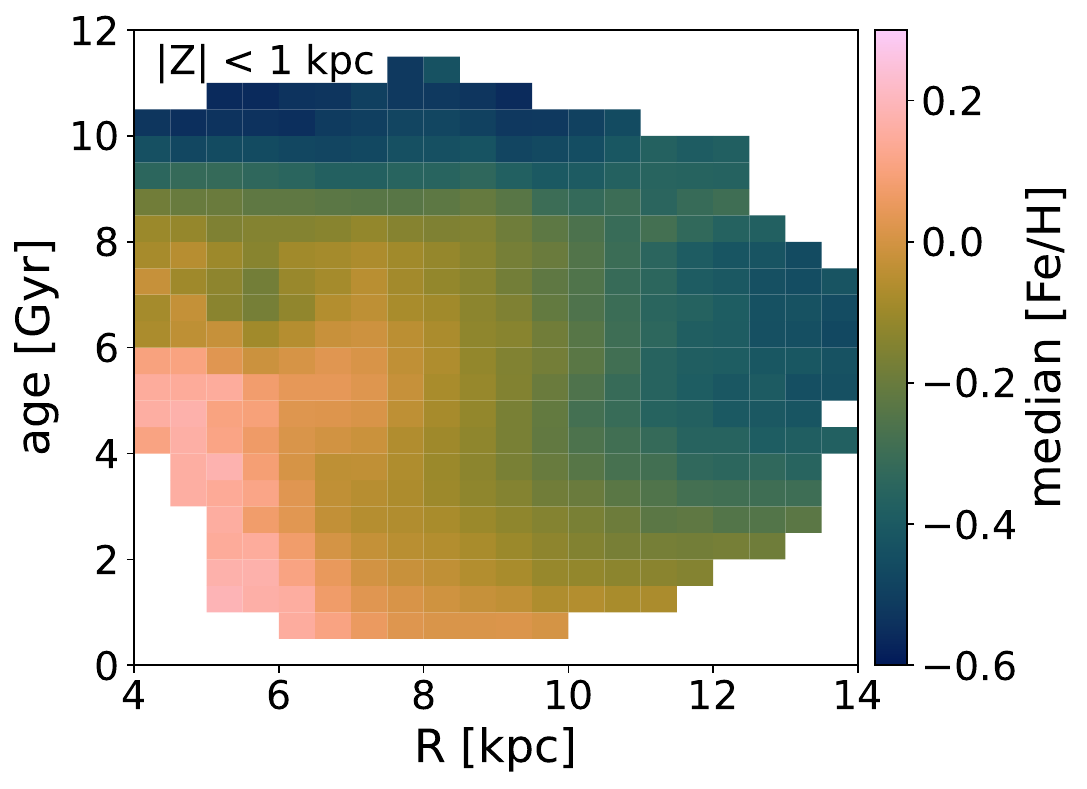}
    \caption{Median [Fe/H] as a function of Galactocentric radius $R$ and age for stars with $|Z|<1$~kpc. The radial gradient at fixed age is the horizontal colour change, which is steepest at intermediate ages.}
    \label{fig:feh_R_age_heatmap}
\end{figure}

We find that the radial metallicity gradient has a strong time-dependence, and that the dependence is non-linear.
For the youngest stars (age $\lesssim2$~Gyr) the gradient is relatively shallow, ${\rm d[Fe/H]/d}R\approx-0.03$~dex/kpc.
It steepens to a maximum of ${\rm d[Fe/H]/d}R\approx-0.07$~dex/kpc at intermediate ages (age $\approx5$--$6$~Gyr), and then flattens sharply for the oldest stars, becoming consistent with zero at age $\gtrsim10$~Gyr.
This ``U-shaped'' age dependence is clearly visible in the flat-steep-flat pattern of the curves in the left panel (as we move across time), as well as the blue curve of the right-panel.
The same pattern is visible in Figure~\ref{fig:feh_R_age_heatmap}, which shows the median [Fe/H] as a function of $R$ and age.

This behaviour is consistent with previous age-resolved measurements.
\citet{xiang2015evolutionstellarmetallicitygradients}, using LAMOST main-sequence turn-off stars, report a steepest radial gradient at intermediate ages that flattens for the oldest populations, with a turning point at ${\sim}7$--$8$~Gyr.
\cite{Anders2017}, using CoRoT--APOGEE red giants, similarly find a flat gradient for the oldest stars that steepens to a maximum before flattening again toward the present day.
\citet{Wang2019}, using LAMOST turn-off and subgiant stars, find the steepest gradient at $4$--$6$~Gyr, and \citet{Anders2023}, using APOGEE giants, find a gradient that is steeper for intermediate-age stars than for young stars and that is consistent with zero at ${>}10$~Gyr.

Studies that instead find a gradient that flattens monotonically with age measure it against guiding radius and select stars on vertically cold orbits \citep{Vickers2021, Willett2023, chen2026}.
In those same data, when the gradient is measured against present-day radius \citep{Vickers2021, Willett2023} or for stars on vertically hotter orbits \citep{Vickers2021, chen2026}, the trend becomes non-monotonic, with the youngest stars shallower than intermediate age stars.
\citet{Wang2019} similarly find that near the midplane, the gradients of stars younger than $8$~Gyr are steeper against present-day radius than against guiding radius, although the intermediate-age steepening of \citet{Anders2023} appears against both.
The shape of the age trend thus appears to depend on the radial coordinate and on how vertically cold the sample is.
Our measurement, against present-day radius over $|Z|<1$~kpc with a purely chemical split, includes the effect of orbital blurring.
\citet{sun2026divergent}, who also use present-day radius and a purely chemical thin disk, find a monotonic flattening, but have no age bin below ${\sim}2$~Gyr.
Our measurements differ from these studies for the youngest stars: our ${\approx}{-}0.03$~dex\,kpc$^{-1}$ below ${\sim}3$~Gyr is roughly half as steep as the ${\approx}{-}0.06$ to $-0.075$~dex\,kpc$^{-1}$ they report for their youngest bins \citep[e.g.,][]{Anders2023, Willett2023, Vickers2021}.
Our thicker vertical selection, with no orbital cut, may contribute, since in those studies the youngest-bin gradient becomes shallower when stars on hotter orbits are included \citep{Vickers2021, Willett2023}, but it is unlikely to be the whole explanation, as \citet{Anders2023} find ${\simeq}{-}0.06$~dex\,kpc$^{-1}$ for their youngest stars within $|Z|<1$~kpc.
The small number and the selection of young giants in our sample (see below) are another possible contributor.

These trends indicate support for particular theories of disk evolution.
Under inside-out growth the inner disk forms and enriches first while the outer disk is still nearly metal-free, so the gas gradient (what the stars are formed from) is expected to be steep at early times.
However, in FIRE-2 simulations the gas gradient instead steepens with time \citep{Graf2024}.
Reconstructions of stellar birth radii, however, infer a gas gradient that was steeper in the past, flattening from ${-}0.15$~dex\,kpc$^{-1}$ at the onset of disk formation \citep{Minchev2018} or about $8$~Gyr ago \citep{Lu2024}, to $-0.07$~dex\,kpc$^{-1}$ for the present day.
Without radial migration, each mono-age population reflects the metallicity gradient in the gas during the time of birth.
The oldest stars should then display the steepest radial gradient, becoming shallower toward the youngest populations.
Indeed, classic inside-out chemical-evolution models without strong radial mixing do not reproduce the old-age gradient that we observe \citep{Chiappini1997}.
Our measurements follow this inside-out growth with no radial migration scenario from the youngest populations up to a maximum near age $\approx5$--$6$~Gyr, but then reverse.
Rather than continuing to steepen, the gradient flattens to near zero for the oldest stars.

This turnover is consistent with efficient radial migration.
Churning near spiral-arm corotation resonances changes stars' guiding radii while largely preserving their near-circular orbits \citep{Sellwood2002}, mixing the disk in two directions (carrying metal-rich stars outward and metal-poor stars inward) so that the median metallicity at each radius regresses toward the global mean, and crucially, flattening the radial metallicity gradient.
The oldest populations, having had the longest to migrate, are the most thoroughly mixed \citep{Roskar2008}.

However, radial migration is not the only possible cause of the flat gradient of the oldest stars.
These stars may also have formed with little radial gradient, in an early, chemically well-mixed disk.
In the FIRE-2 simulations, mono-age gradients today are only modestly shallower than at birth, and the oldest stars have negligible gradients at both times \citep{Graf2024}.
\citet{Minchev2018} also note that an initially flat and an initially negative gas gradient are hard to tell apart for the oldest stars.

In terms of radial migration, observed U-shape could be explained as the result of the tension between two processes that have opposite effects with age: the birth gradient steepens with lookback time due to inside-out growth and the gradient flattening due to radial migration also increases with age.
This combined effect, a steep intrinsic gradient progressively flattened by mixing, peaks at intermediate ages.
This steepen-then-flatten shape is reproduced by chemodynamical models that combine inside-out growth with radial migration \citep{Minchev2014, Kubryk2015}.
Its amplitude also ties into constraints on stellar feedback; in the simulations of \citet{Gibson2013}, weak or conventional feedback produces steep early gradients that flatten with time, whereas enhanced feedback yields flat, nearly time-invariant gradients at all ages.
The steep intermediate-age gradient we measure shows that the disk was not always flat, disfavouring the flattest enhanced-feedback predictions.
We caution that the shallow gradient of the very youngest bin may be affected in part by the small number and possible selection biases of young giants in our sample.

\begin{figure}
    \centering
    \includegraphics[width=0.95\linewidth]{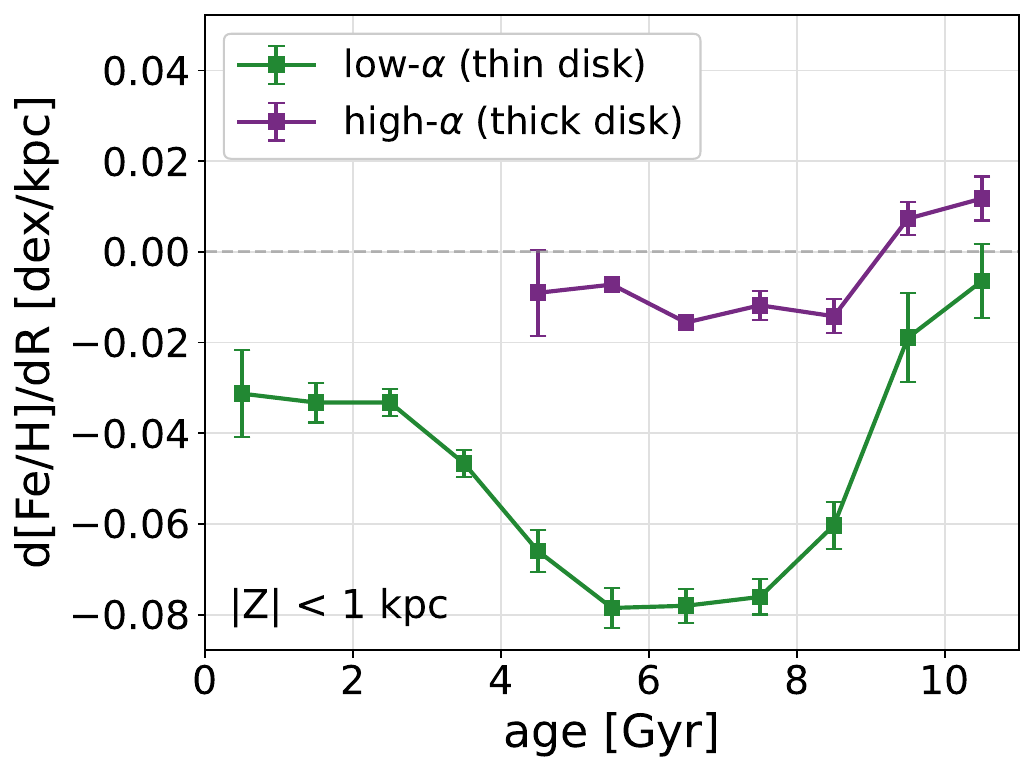}
    \caption{Radial metallicity gradient as a function of age for the low-$\alpha$ (green) and high-$\alpha$ (purple) sequences within $|Z|<1$~kpc. The high-$\alpha$ gradient is weak at all ages, meaning that the U-shaped age dependence of the radial gradient in the full sample is driven by the low-$\alpha$ disk.}
    \label{fig:radial_gradient_thin_thick}
\end{figure}

We also analyze whether the two chemically-distinct sequences differ in their age dependence.
Figure~\ref{fig:radial_gradient_thin_thick} shows the radial gradient as a function of age, measured separately for the low-$\alpha$ (thin-disk) and high-$\alpha$ (thick-disk) sequences of Figure~\ref{fig:mgfe_plane}, for stars within $|Z|<1$~kpc.
The high-$\alpha$ population has a weak radial gradient at all ages, as expected for an old, rapidly-formed, well-mixed component.
Within $|Z|<1$~kpc we find ${\rm d[Fe/H]/d}R=-0.0156\pm0.0013$~dex\,kpc$^{-1}$ for high-$\alpha$ stars aged $6$--$7$~Gyr and ${+}0.0117\pm0.0049$~dex\,kpc$^{-1}$ for those aged $10$--$11$~Gyr (Table~\ref{tab:radial_gradients_age}).
The gradient therefore changes sign across the high-$\alpha$ population, although its amplitude never exceeds ${\sim}0.016$~dex\,kpc$^{-1}$, significantly weaker than the gradients observed in the low-$\alpha$ sequence.
The U-shaped age dependence of the full sample is therefore largely driven by the low-$\alpha$ population, which reaches ${\rm d[Fe/H]/d}R\approx-0.08$~dex/kpc at intermediate ages.
A flat thick-disk gradient alongside a steeper, age-dependent thin-disk gradient is consistent with the results of \citet{sun2024mappingchemodynamicsgalacticdisk}.
The mildly positive gradient we measure for the oldest high-$\alpha$ stars also agrees with recent work targeting the thick disk.
\citet{chen2026}, using LAMOST DR8 with asteroseismically calibrated ages, report a positive thick-disk gradient of ${\sim}{+}0.013$~dex\,kpc$^{-1}$ that is roughly constant with age, and \citet{sun2026divergent}, using LAMOST and APOGEE red clump stars, find a weakly positive thick-disk gradient that is invariant with vertical height and age. 
In the $10$--$11$~Gyr bin, the ${+}0.0117\pm0.0049$~dex\,kpc$^{-1}$ that we measure is consistent with both, although the slight negative gradient we measure for younger stars is not.

\subsubsection{Face-on maps of the radial gradient residuals}
\label{sec:feh_residual_maps}

\begin{figure*}[th]
    \centering
    \includegraphics[width=0.9\linewidth]{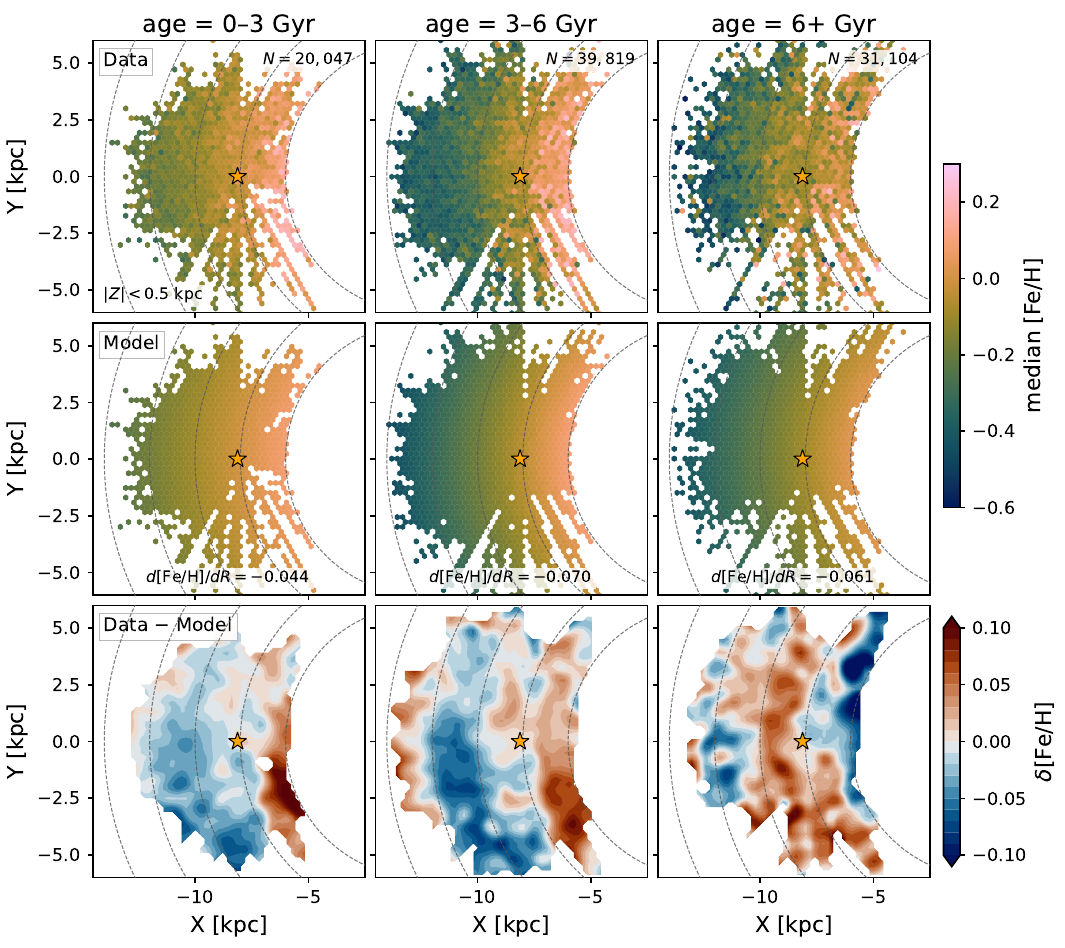}
    \caption{Maps of [Fe/H] in the X-Y plane for APOGEE giants at $|Z|<0.5$~kpc, in three bins of stellar age (columns). \textbf{Top}: median [Fe/H] in hexbin cells that contain at least three stars. \textbf{Middle}: the linear radial-gradient fit of [Fe/H] to each age bin, with the fitted slope in dex\,kpc$^{-1}$. \textbf{Bottom}: data minus model, computed as the median residual in $0.2$~kpc cells and smoothed with a Gaussian of $\sigma=0.25$~kpc. The star marks the Sun, and the dashed circles are at $R=6$, $8$, $10$, $12$, and $14$~kpc.}
    \label{fig:feh_residual_maps}
\end{figure*}

Previous work, such as \cite{Hackshaw2024}, has gone beyond measuring the radial gradient to examine the residuals of the fit, investigating whether there is azimuthal structure in the disk.
We follow a similar approach, including the age dimension.
In Figure \ref{fig:feh_residual_maps} we show face-on (X-Y) maps of the median [Fe/H] in three age bins (top row), the linear radial-gradient fit to each bin (middle row), and the residuals of the fit (bottom row).
The Sun's location is marked with a star, and the dashed circles are at $R=6$, $8$, $10$, $12$, and $14$~kpc.
We build these maps from APOGEE alone, the only surveys that samples the plane far from the Sun sufficiently well.
We select giants with $\logg < 3.6$, $3500 < \teff < 5000$~K, fractional distance uncertainty below $30\%$, $|Z|<0.5$~kpc, and $6<R<14$~kpc.
We fit the stars in each age bin with a line through the median [Fe/H] of $1$~kpc radial bins, and smooth the residuals with a Gaussian of $\sigma=0.25$~kpc.

We find that the residuals show a clear oscillatory pattern in radius. 
Interestingly, the pattern is inverted for the oldest stars as compared to the younger stars.
In particular, for the two younger age bins, around $R \approx 7$~kpc the median [Fe/H] lies above the linear fit (i.e., the bins are red), whereas for $R \approx 10$~kpc it lies below (i.e., the bins are blue).
On the other hand, for the oldest age bin, the median [Fe/H] lies below the linear fit at $R \approx 7$~kpc and above it at $R \approx 10$~kpc.

\subsection{Age-resolved vertical metallicity gradients}

\begin{figure*}[th]
    \centering
    \includegraphics[width=1\linewidth]{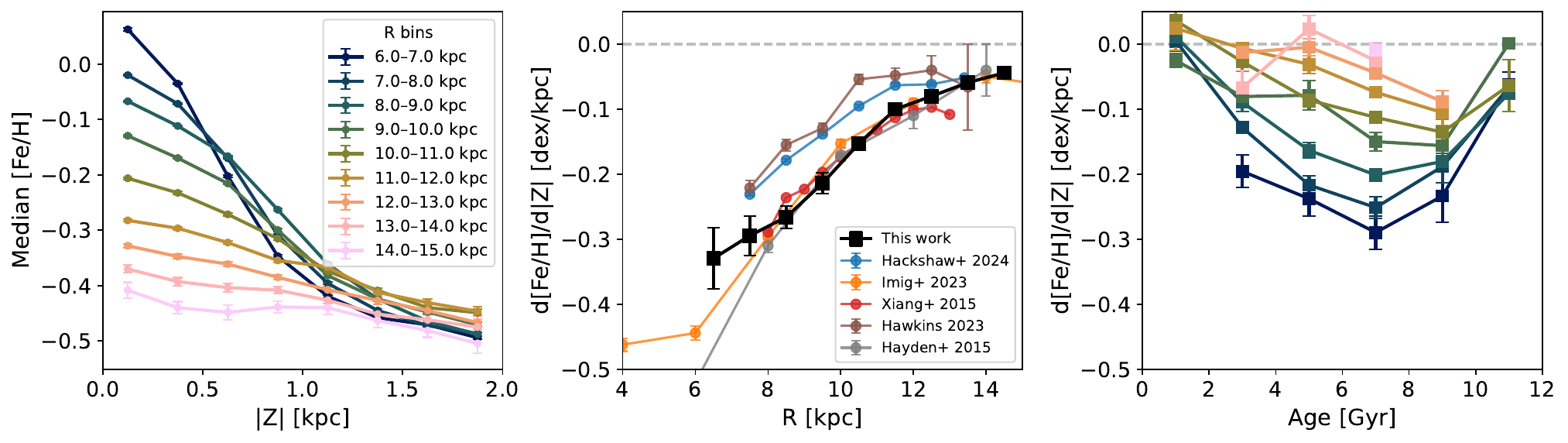}
    \caption{Vertical metallicity profile and gradient. \textbf{Left}: median [Fe/H] versus $|Z|$ in bins of $R$. \textbf{Middle}: the vertical gradient as a function of $R$ (black), compared to literature measurements. \textbf{Right}: the vertical gradient as a function of age, in bins of $R$.}
    \label{fig:vertical_metallicity_profile}
\end{figure*}

In Figure \ref{fig:vertical_metallicity_profile} we show the vertical metallicity structure of the Galaxy.
The left panel shows the median [Fe/H] as a function of vertical distance $|Z|$ from the Galactic midplane, in bins of Galactocentric radius $R$.
Here we have required that each bin contains at least 50 samples in order to be included.
Error bars denote the standard error on the median in each $|Z|$ bin. 
In the inner regions of the Galaxy (i.e., small $R$), [Fe/H] decreases steeply with increasing $|Z|$, reflecting the sharp transition from the metal-rich thin disk to the metal-poor thick disk.
As we move further out to larger $R$, [Fe/H] decreases only a very small amount with increasing $|Z|$, reflecting greater homogeneity between the thin and thick disks.
The sharpness of the transition can also be seen in Figure \ref{fig:spatial_distribution}, where at large $R$ the distinction between the two regions is blurrier.
The same qualitative trends are also observed by, for example, \citet{Imig_2023} and \citet{gaia2023chemicalcartography}.

We quantify this steepness in the change of [Fe/H] with respect to $|Z|$ in the middle panel of Figure \ref{fig:vertical_metallicity_profile} by performing linear fits to the profiles in the left panel.
We measure that in the $6{<}R{<}7~$kpc bin, the gradient is ${\rm d[Fe/H]/d|Z|}=-0.33\pm0.05~$dex/kpc, which shallows out eventually to ${\rm d[Fe/H]/d|Z|}=-0.080\pm0.002~$dex/kpc in the $12{<}R{<}13~$kpc bin.
In the solar neighbourhood, using a sample-count weighted average of our results in the $7{<}R{<}8$ and $8{<}R{<}9$ bins, we find that ${\rm d[Fe/H]/d|Z|}=-0.28\pm0.02$.
Beyond the solar neighbourhood, the change in the gradient is roughly linear, and for $R<8~$kpc, it also looks linear but with a shallower slope.
Our exact results are in Table \ref{tab:vertical_gradients}.

\begin{table}
  \centering
  \caption{Vertical metallicity gradients as a function of $R$.}
  \label{tab:vertical_gradients}
  \begin{tabular}{cccc}
    \hline
    $R_{\rm min}$ [kpc] & $R_{\rm max}$ [kpc] & $N$ & d[Fe/H]/d$|Z|$ [dex/kpc] \\
    \hline

    6.00 & 7.00 & 81123 & $-0.3294 \pm 0.0469$ \\
    7.00 & 8.00 & 178271 & $-0.2944 \pm 0.0306$ \\
    8.00 & 9.00 & 144964 & $-0.2658 \pm 0.0173$ \\
    9.00 & 10.00 & 64599 & $-0.2138 \pm 0.0160$ \\
    10.00 & 11.00 & 32787 & $-0.1528 \pm 0.0088$ \\
    11.00 & 12.00 & 17710 & $-0.1001 \pm 0.0047$ \\
    12.00 & 13.00 & 8309 & $-0.0800 \pm 0.0015$ \\
    13.00 & 14.00 & 3010 & $-0.0590 \pm 0.0032$ \\
    14.00 & 15.00 & 945 & $-0.0441 \pm 0.0081$ \\
    \hline
  \end{tabular}
\end{table}

We compare our results to a number of literature measurements in the middle panel of Figure \ref{fig:vertical_metallicity_profile}.
In the solar neighbourhood,
\citet{Hayden2015} find that ${\rm d[Fe/H]/d|Z|}=-0.31\pm0.01$ in a $7~{\rm kpc}<R<9~{\rm kpc}$ bin, and
\citet{Imig_2023} measure
${\rm d[Fe/H]/d|Z|}=-0.296\pm0.009$ at $R=8~$kpc.
More broadly, our results are highly consistent with the measurements from \citet{Imig_2023}, \citet{Hayden2015}, and \citet{xiang2015evolutionstellarmetallicitygradients} across the Galactocentric radius range of $8~{\rm kpc}<R<12~{\rm kpc}$. 
\citet{Hackshaw2024} and \citet{Hawkins_2023}, in this region, measure slightly more shallow vertical gradients, but all the literature measurements seem to converge to around ${\rm d[Fe/H]/d|Z|}=-0.05~$dex/kpc around $R=13~$kpc, in good agreement with our measurements.

Given that we have in hand stellar ages, we can also trace how the vertical metallicity gradient varies with age.
The right panel of Figure \ref{fig:vertical_metallicity_profile} shows, once again in different Galactocentric radius bins, the vertical metallicity gradient as a function of age.
All measurements again require more than 50 samples per bin and at least 3 valid bins for the linear fit.
We find distinct U-shaped curves, with small gradients at the present and far in the past and larger gradients at $6{-}8~$Gyr.
Additionally, the dependence of the gradient on $R$ is also greater at $6{-}8~$Gyr, whereas the vertical gradient is essentially the same at all radii at the present and far in the past.

This dependence of the vertical metallicity gradient on age has also been studied in the past by a few works.
\citet{Ciuca_2017}, using a sample of $N\approx 18{,}000$ dwarf stars from a cross-match between Tycho-Gaia Astrometric Solution (TGAS) and RAdial Velocity Experiment (RAVE), find that for young stars, the vertical metallicity gradient is consistent with zero.
For older stars they find that it is negative, with gradients as strong as ${\rm d[Fe/H]/d|Z|}=-0.65\pm0.12$ or ${\rm d[Fe/H]/d|Z|}=-0.43\pm0.12$ (with and without accounting for stellar population bias correction, respectively) for the $8-11~$Gyr age bin.
However, their sample is very limited spatially, covering only the solar neighbourhood at $|Z|{<}0.5~$kpc and $7.6{<}R{<}8.4~$kpc, with no old stars past $11~$Gyr.

\citet{long2025verticalmetallicitygradientgalactic}, using a young stellar sample from LAMOST and open clusters, find, consistent with \citet{Ciuca_2017}, that within a mono-age population for young stars (their sample is limited to stars with ages of $<3~$Gyr), the vertical metallicity gradient is consistent with zero for all age bins.

\citet{xiang2015evolutionstellarmetallicitygradients} use a large sample ($N\approx 300{,}000$) of LAMOST MSTO stars and ages from isochrone fitting to measure the evolution of metallicity gradients.
They find that for the oldest stars, there is a negative gradient of ${\rm d[Fe/H]/d|Z|}=-0.11\pm0.02~$dex/kpc with minimal dependence on $R$.
The steepest gradient is found for stars in their $8{-}11~$Gyr and $6{-}8~$Gyr bins,
with once again a shallow gradient for young stars between $2{-}4~$Gyr.
However, they have no young stars with ages ${<}2~$Gyr in their sample.
Regardless, in the regime where they do have samples, their results agree very well with ours;
they note a turning point in the gradients at ${\sim}8~$Gyr, which we also observe.
With their sample they could not conclude whether this turning point was dependent on $R$.
     
The existing literature paints a consistent picture, which we unify here.
For the youngest stars, with ages less than ${\sim}2~$Gyr, the vertical metallicity gradient is close to zero across most of the disk, with only a weak $R$-dependence.
On the other end, for the oldest stars with ages greater than ${\sim}10~$Gyr, the gradient is slightly negative, on the order of ${\rm d[Fe/H]/d|Z|}=-0.06~$dex/kpc, again with minimal dependence on $R$.
However, for intermediate-age stars, the relation ``fans out:'' the U-shape is more pronounced for stars closer to the Galactic center.
In other words, for stars at smaller $R$, the gradient at intermediate ages is more strongly negative than for stars at larger $R$.

These vertical gradient trends constrain how the disk was assembled vertically.
In the ``upside-down'' formation picture seen in cosmological simulations \citep{Brook2004, Bird2013}, the disk forms in a thick, turbulent, gas-rich state and settles into progressively thinner layers as it grows, so the oldest stars are born kinematically hot and vertically extended rather than being a thin population heated only later.
This can explain the vertically-extended, metal-poor old populations and the concentration of the steep intermediate-age vertical gradient toward the inner Galaxy, where star formation settled earliest.
Secular heating of an initially-thin disk is also a factor but may be too weak to explain the hottest, oldest stars on its own \citep[e.g.,][]{Frankel2020};
the early accretion of \textit{Gaia}-Enceladus \citep{Helmi2018} likely supplied additional heating at early times.

\subsection{[Mg/Fe] profiles}
\label{sec:mgfe_gradient}

The [Fe/H] gradient traces how the disk built up its total metallicity content, but it does not distinguish between two enrichment channels with different timescales.
Given the different timescales of the two enrichment channels, the ratio [$\alpha$/Fe] is a proxy for local star-formation timescale, rather than the accumulated amount of enrichment.
In line with many other works, we use [Mg/Fe] as our $\alpha$ tracer; magnesium is a nearly pure core-collapse product and a reliably measured element in our catalog.
The other $\alpha$ elements (Si, Ca, Ti) give fairly consistent results to what we show next, and they are shown in Appendix~\ref{apdx:alpha}.
Contrasting the radial gradients of [Mg/Fe] and [Fe/H] as a function of age (Figure~\ref{fig:radial_gradient_vs_age}, middle and right panels) thus distinguishes how fast stars formed from how much metal had accumulated at a given place and time.

We observe that the two behave very differently.
Whereas the [Fe/H] gradient is steep and strongly age-dependent (the U-shape of Figure~\ref{fig:radial_gradient_vs_age}), the [Mg/Fe] gradient is weak at all ages, with $|{\rm d[Mg/Fe]}/{\rm d}R| \lesssim 0.01$~dex\,kpc$^{-1}$---several times shallower than iron and shows no resemblance to the U-shape.
The mono-age [Mg/Fe] profiles (Figure~\ref{fig:radial_gradient_vs_age}, middle) are nearly flat with radius (also seen in [Ti/Fe], see Figure \ref{fig:tife_gradient}), and are essentially just ordered in amplitude by age, with the oldest populations being the most $\alpha$-enhanced (which is also observed in Figure~\ref{fig:mgfe_plane}).
At intermediate ages, the [Mg/Fe] gradient is slightly positive (${\rm d[Mg/Fe]}/{\rm d}R \approx 0.01$~dex\,kpc$^{-1}$), and at old ages, it is slightly negative (${\rm d[Mg/Fe]}/{\rm d}R \approx -0.003$~dex\,kpc$^{-1}$).
This is also observed in the [Si/Fe] and [Ca/Fe] gradients in Figures \ref{fig:sife_gradient} and \ref{fig:cafe_gradient}.
In any case, the weak radial [Mg/Fe] gradient, close to zero, is consistent with previous RAVE measurements near the midplane: \citet{Boeche2013dwarf} find ${\rm d[Mg/Fe]}/{\rm d}R_g = -0.009\pm0.002$~dex\,kpc$^{-1}$ for dwarfs (measured against guiding radius $R_g$), and \citet{Boeche2014giant} find ${\rm d[Mg/Fe]}/{\rm d}R = +0.020\pm0.004$~dex\,kpc$^{-1}$ for giants.

Physically, a nearly flat [Mg/Fe] gradient would be expected if the balance between fast and delayed enrichment was close to uniform across the disk at every epoch.
This is what the two-phase picture predicts: the old, $\alpha$-rich population formed rapidly and disk-wide during an early, vigorous star-forming phase, imprinting little radial variation in [Mg/Fe] \citep{Haywood2013, Nidever2014}, while the strong [Fe/H] gradient was established later by inside-out growth and subsequently reshaped by radial migration \citep{Chiappini1997, Sellwood2002}.

We note that the focus of this work is the chemical structure of the disk and its evolution, not its kinematics. 
A dynamical view of the same two-phase picture---the age--velocity-dispersion relation, and the disk heating it encodes---requires radial velocities, which are not part of the catalog of \citet{Shen2026} and which we have not compiled for this sample, and we thus defer it to future work, along with a full chronochemodynamical analysis.

\subsection{Age-metallicity relation}

\begin{figure}
    \centering
    \includegraphics[width=0.95\linewidth]{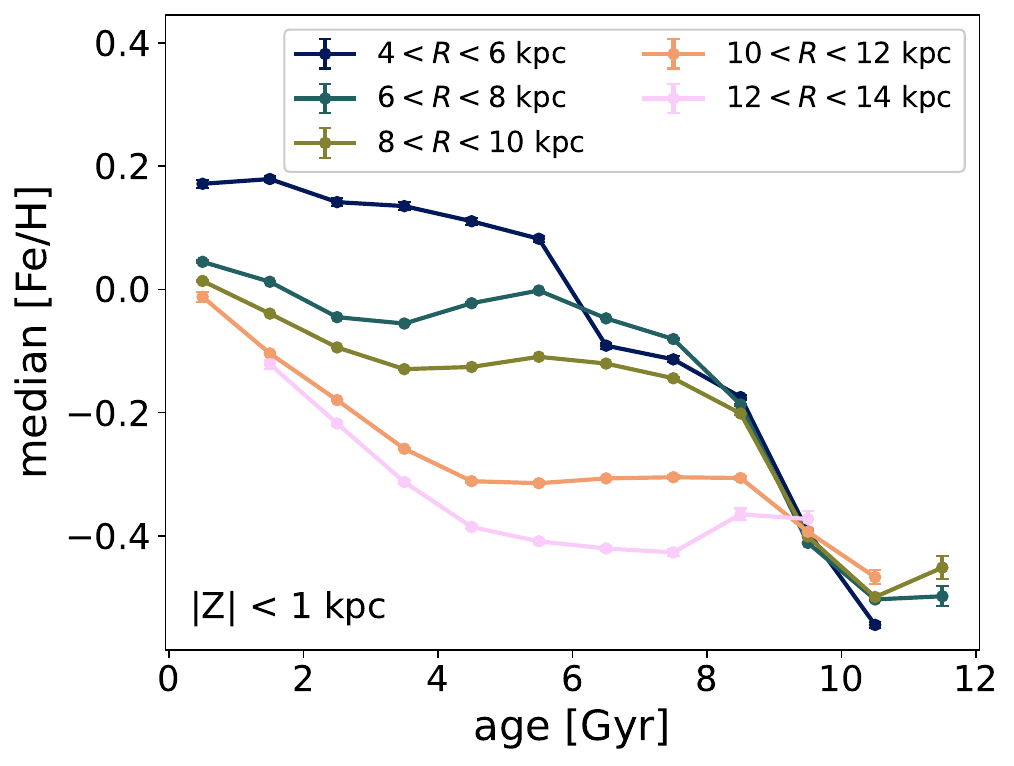}
    \caption{Age--metallicity relation in bins of Galactocentric radius for $|Z|<1$~kpc stars. Median [Fe/H] declines with age at all radii. The radial spread between the curves (i.e., the radial metallicity gradient) is largest at intermediate ages and disappears for the oldest stars, which converge to $\feh\approx-0.45$ by age $\approx10$~Gyr for all radii.}
    \label{fig:age_metallicity_by_R}
\end{figure}

Since the age dependence of the metallicity gradient is the spatial derivative of the age--metallicity relation (AMR), it is also useful to analyze the AMR directly.
Figure~\ref{fig:age_metallicity_by_R} shows the median [Fe/H] as a function of age in bins of Galactocentric radius for $|Z|<1$~kpc.
The inner disk ($4<R<6$~kpc) is the most metal-rich, starting off at [Fe/H]$\approx 0.17$~dex at age $<2$~Gyr and declining with age, whereas the outer disk is metal-poor throughout.
The radial spread between the curves, which represents the radial gradient, is largest at intermediate ages ($\sim 5$~Gyr) but converges for older stars.
Indeed, by age $\approx 10$~Gyr, stars at all radii have [Fe/H]$\approx -0.45$~dex with minimal spread across radii.
This signal that we observe in the AMR is the same signal as in Figures~\ref{fig:radial_gradient_vs_age} and \ref{fig:feh_R_age_heatmap}, but seen from a complementary angle.
The overall decline of [Fe/H] with age and its spatial (radial) dependence are consistent with the observations in \citet{Feuillet2019}, and the large scatter of the AMR at any single radius, especially in the solar neighbourhood, is a well-known consequence of radial migration \citep{Sellwood2002} bringing together stars of different birth radii (and hence different metallicities) to a common present-day location.

\section{Conclusion}
\label{sec:conclusion}

We have used the catalog of \cite{Shen2026}---$581{,}688$ giants with uniform stellar parameters, abundances, distances, and ages derived from five spectroscopic surveys in a self-consistent manner---to map the metallicity structure of the Milky Way and to trace how it has evolved over time.
Our main findings are as follows:

\begin{itemize}
    \item The radial [Fe/H] gradient varies non-monotonically with age. It is shallow for the youngest stars (${\rm d[Fe/H]/d}R\approx-0.03$~dex\,kpc$^{-1}$ at age $\lesssim2$~Gyr), steepens to a maximum of ${\sim}{-}0.07$~dex\,kpc$^{-1}$ at intermediate ages (age $\approx5$--$6$~Gyr), and flattens to near zero for the oldest populations (age $\gtrsim10$~Gyr), reaching a slightly positive ${+}0.012\pm0.006$~dex\,kpc$^{-1}$ at $10$--$11$~Gyr. This ``U-shaped'' age dependence, which is strongly $|Z|$-dependent, agrees with previous age-resolved measurements that use present-day radius.
    \item The U-shape is driven almost entirely by the thin disk. The high-$\alpha$ (thick-disk) sequence has a weak radial gradient at all ages, whereas the low-$\alpha$ (thin-disk) sequence has strong age dependence, reaching ${\rm d[Fe/H]/d}R\approx-0.08$~dex\,kpc$^{-1}$ at intermediate ages.
    \item The vertical [Fe/H] gradient shows a similar U-shaped age dependence, which is strongly dependent on $R$. At $0<|Z|<2$~kpc, it steepens from ${\rm d[Fe/H]/d}|Z|=-0.080\pm0.002$~dex\,kpc$^{-1}$ in the outer disk ($12<R<13$~kpc) to $-0.33\pm0.05$~dex\,kpc$^{-1}$ in the inner disk ($6<R<7$~kpc). For the youngest stars the vertical gradient is close to zero, for the oldest stars the gradient is ${\sim}{-}0.06$~dex\,kpc$^{-1}$, and at intermediate ages the gradient is steepest, and especially so in the inner disk, reaching ${\sim}{-}0.3$~dex\,kpc$^{-1}$.
    \item The radial [Mg/Fe] gradient is weak at all ages. At every epoch, we measure $|{\rm d[Mg/Fe]}/{\rm d}R|\lesssim0.01$~dex\,kpc$^{-1}$---several times shallower than for [Fe/H] and showing no U-shape; other $\alpha$ elements (Si, Ca, Ti) behave similarly ($|{\rm d}[X/{\rm Fe}]/{\rm d}R|\lesssim0.01$~dex\,kpc$^{-1}$).
    \item The age--metallicity relation shows a large radial spread in median [Fe/H] at intermediate ages, and converges to $\feh\approx-0.45$ by age $\approx10$~Gyr, independent of radius.
\end{itemize}

Taken together with the age-resolved [Mg/Fe]--[Fe/H] plane and the age--metallicity relation, these results support a two-phase picture: an early, vigorous, chemically well-mixed phase that formed the old, $\alpha$-rich component with little radial structure, followed by extended inside-out growth of a thin disk in which the radial gradient of the star-forming gas has flattened over at least the last ${\sim}6$~Gyr; the near-zero gradient of the oldest stars reflects radial migration erasing the gradient they were born with, their formation in the early, well-mixed phase before a gradient was established, or both.
The weak [Mg/Fe] gradient indicates that this radial structure was set by the spatial build-up of iron.

The homogeneity of the underlying catalog was central to this analysis, removing the cross-survey abundance systematics that could otherwise masquerade as spatial gradients.

\subsection{Limitations and Future Work}

We inherit all of the limitations of the catalog of \cite{Shen2026}.
In particular, while cross-survey systematics are eliminated, we inherit all the biases of the underlying label set, here AstroNN abundances and APOKASC ages \citep{Leung2018}.
Our sample only probes stars with a limited metallicity range; in particular, we have essentially no stars with [Fe/H]${<}-0.8$~dex, limiting our analysis in the metal-poor regime.
Furthermore, our age estimates are reliable only for giants, and thus only a small subset of the released catalog is usable for our analysis.
Our age and distance estimates carry non-negligible per-star uncertainties.
Further, the ages inherit a compression of the youngest and oldest ages from the training labels, which \citet{Shen2026} show (based on calibrations on open clusters) but do not quantify.
However, we have tested whether these propagate into the measured trends by performing a Monte Carlo resampling of the distances, ages, and [Fe/H] values within their uncertainties and recomputing the gradient evolution (the shaded band in Figure~\ref{fig:radial_gradient_vs_age}).
The resulting $16$--$84$th percentile band is much narrower than the amplitude of the trend, so the curve is stable to added noise at the level of the catalog uncertainties.
This test does not address age errors that are larger than the catalog uncertainties (median $0.2$--$0.4$~Gyr) or systematic, which we examine with an injection test in Appendix~\ref{apdx:ages}; age errors that correlate with [Fe/H] or [$\alpha$/Fe] were not tested.

Selection effects are also a concern in our analysis.
The five contributing surveys have spatially non-uniform selection functions that are far too complicated for us to model in this work.
Because our labels are homogeneous, cross-survey systematics cannot imprint spurious abundance gradients (Section~\ref{sec:data}), but spatial sampling biases can.
Recomputing the age dependence independently within each survey recovers the same steepen-then-flatten trend in every survey that spans the relevant age range (Appendix~\ref{apdx:selection}), which rules out a bias unique to any single survey, but does not eliminate any common biases.
The chief shared bias is the magnitude limit, which at large distances preferentially selects the most luminous giants, and we do not model it.
If such giants are, at fixed age, modestly more metal-poor, the magnitude limit would bias the absolute normalization of the gradient at large $R$.
It would not, however, produce the age dependence we report: the effect would steepen the negative gradient and reduce outer-disk metallicities most strongly for the oldest populations, working against, rather than creating the flat gradient we measure at old ages.

Taking the framework of \citet{Shen2026} and deriving homogeneous radial velocities for this subset of stars, producing 6D phase space information, would be highly complementary to the chemical analysis we have done here and allow for a full chronochemodynamical analysis of the Galaxy.
The resulting kinematic information would also allow the gradients to be measured against guiding radius and orbital actions and would enable a cleaner dynamical separation of the disk components.
Furthermore, extending our analysis here to the full suite of ${\sim}20$ elements in the catalog would allow us to trace the disk's chemical evolution in far more detail.
We defer both extensions to future work.

\begin{acknowledgments}

J.S. is supported by the Natural Sciences and Engineering Research Council of Canada (NSERC), funding reference number 587652.
J.S.S. was supported by funding from NSERC Discovery Grant RGPIN-2023-04849.
The computations in this work were carried out on, and the data hosted on equipment supported by the Scientific Computing Core at the Flatiron Institute, a division of the Simons Foundation.
We would like to acknowledge the support of the Simons Foundation and of Schmidt Sciences. 
This work was supported in part by the AI2050 program at Schmidt Sciences (Grant G-25-70028).

This work made use of the Third Data Release of the GALAH Survey (Buder et al. 2021). The GALAH Survey is based on data acquired through the Australian Astronomical Observatory, under programs: A/2013B/13 (The GALAH pilot survey); A/2014A/25, A/2015A/19, A2017A/18 (The GALAH survey phase 1); A2018A/18 (Open clusters with HERMES); A2019A/1 (Hierarchical star formation in Ori OB1); A2019A/15 (The GALAH survey phase 2); A/2015B/19, A/2016A/22, A/2016B/10, A/2017B/16, A/2018B/15 (The HERMES-TESS program); and A/2015A/3, A/2015B/1, A/2015B/19, A/2016A/22, A/2016B/12, A/2017A/14 (The HERMES K2-follow-up program). We acknowledge the traditional owners of the land on which the AAT stands, the Gamilaraay people, and pay our respects to elders past and present. This paper includes data that has been provided by AAO Data Central (datacentral.org.au).

This work has made use of data from the European Space Agency (ESA) mission
{\it Gaia} (\url{https://www.cosmos.esa.int/gaia}), processed by the {\it Gaia}
Data Processing and Analysis Consortium (DPAC,
\url{https://www.cosmos.esa.int/web/gaia/dpac/consortium}). Funding for the DPAC
has been provided by national institutions, in particular the institutions
participating in the {\it Gaia} Multilateral Agreement.

This research used data obtained with the Dark Energy Spectroscopic Instrument (DESI). DESI construction and operations is managed by the Lawrence Berkeley National Laboratory. This material is based upon work supported by the U.S. Department of Energy, Office of Science, Office of High-Energy Physics, under Contract No. DE–AC02–05CH11231, and by the National Energy Research Scientific Computing Center, a DOE Office of Science User Facility under the same contract. Additional support for DESI was provided by the U.S. National Science Foundation (NSF), Division of Astronomical Sciences under Contract No. AST-0950945 to the NSF’s National Optical-Infrared Astronomy Research Laboratory; the Science and Technology Facilities Council of the United Kingdom; the Gordon and Betty Moore Foundation; the Heising-Simons Foundation; the French Alternative Energies and Atomic Energy Commission (CEA); the National Council of Humanities, Science and Technology of Mexico (CONAHCYT); the Ministry of Science and Innovation of Spain (MICINN), and by the DESI Member Institutions: www.desi.lbl.gov/collaborating-institutions. The DESI collaboration is honored to be permitted to conduct scientific research on I’oligam Du’ag (Kitt Peak), a mountain with particular significance to the Tohono O’odham Nation. Any opinions, findings, and conclusions or recommendations expressed in this material are those of the author(s) and do not necessarily reflect the views of the U.S. National Science Foundation, the U.S. Department of Energy, or any of the listed funding agencies.

Funding for the Sloan Digital Sky 
Survey IV has been provided by the 
Alfred P. Sloan Foundation, the U.S. 
Department of Energy Office of 
Science, and the Participating 
Institutions. 

SDSS-IV acknowledges support and 
resources from the Center for High 
Performance Computing at the 
University of Utah. The SDSS 
website is www.sdss4.org.

SDSS-IV is managed by the 
Astrophysical Research Consortium 
for the Participating Institutions 
of the SDSS Collaboration including 
the Brazilian Participation Group, 
the Carnegie Institution for Science, 
Carnegie Mellon University, Center for 
Astrophysics | Harvard \& 
Smithsonian, the Chilean Participation 
Group, the French Participation Group, 
Instituto de Astrof\'isica de 
Canarias, The Johns Hopkins 
University, Kavli Institute for the 
Physics and Mathematics of the 
Universe (IPMU) / University of 
Tokyo, the Korean Participation Group, 
Lawrence Berkeley National Laboratory, 
Leibniz Institut f\"ur Astrophysik 
Potsdam (AIP),  Max-Planck-Institut 
f\"ur Astronomie (MPIA Heidelberg), 
Max-Planck-Institut f\"ur 
Astrophysik (MPA Garching), 
Max-Planck-Institut f\"ur 
Extraterrestrische Physik (MPE), 
National Astronomical Observatories of 
China, New Mexico State University, 
New York University, University of 
Notre Dame, Observat\'ario 
Nacional / MCTI, The Ohio State 
University, Pennsylvania State 
University, Shanghai 
Astronomical Observatory, United 
Kingdom Participation Group, 
Universidad Nacional Aut\'onoma 
de M\'exico, University of Arizona, 
University of Colorado Boulder, 
University of Oxford, University of 
Portsmouth, University of Utah, 
University of Virginia, University 
of Washington, University of 
Wisconsin, Vanderbilt University, 
and Yale University. 

\end{acknowledgments}

\software{Astropy \citep{astropy:2013, astropy:2018, astropy:2022}, Numpy \citep{Harris2020numpy}, Matplotlib \citep{Hunter:2007}, Pandas \citep{mckinney-proc-scipy-2010, reback2020pandas}, SciPy \citep{2020SciPy}}

\appendix

\section{Robustness to survey selection}
\label{apdx:selection}

Our sample is drawn from five spectroscopic surveys with very different (and complex) selection functions and footprints (see Figure~\ref{fig:coverage_by_survey}).
For example, APOGEE and DESI observe in pencil beams, while \textit{Gaia} RVS samples the solar neighbourhood almost uniformly.
Because we do not model the combined selection function (Section~\ref{sec:data}), which is far too complicated, we instead test whether our result on the U-shaped age dependence of the radial [Fe/H] gradient is sensitive to individual selection functions of any particular survey by recomputing the relation independently for each survey.
If the U-shaped evolution is an artifact of how a particular survey samples the disk, the per-survey curves would disagree.
In contrast, a real signal should appear in all of them.

\begin{figure}
    \centering
    \includegraphics[width=0.98\linewidth]{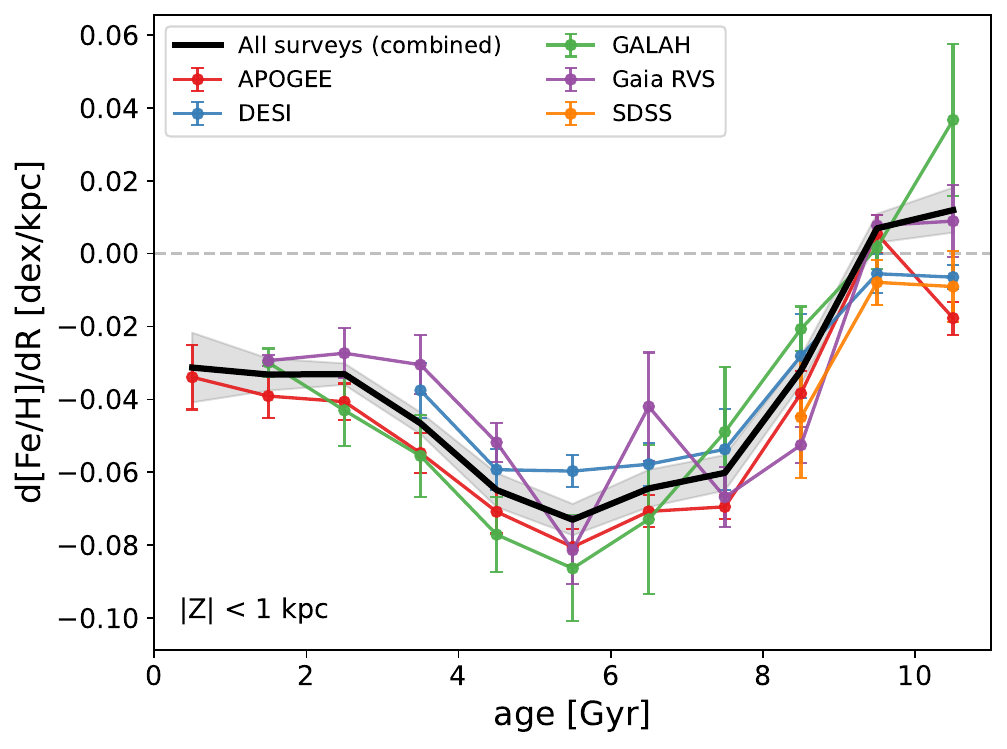}
    \caption{Radial [Fe/H] gradient as a function of stellar age, recomputed independently for each of the five contributing surveys (coloured points) and for the combined sample (black line). The shaded band is the linear-fit uncertainty. All curves use the identical binning and fitting of Figure~\ref{fig:radial_gradient_vs_age}, and a survey is shown in an age bin only where it has enough stars to constrain the fit (${>}50$ per radial bin, $\geq3$ radial bins). Despite their different selection functions and footprints, all surveys with the relevant age coverage---APOGEE, GALAH, \textit{Gaia} RVS, and DESI---independently recover the same U-shaped evolution; SDSS appears in only three old-age bins and does not constrain the shape.}
    \label{fig:robustness_per_survey}
\end{figure}

Figure~\ref{fig:robustness_per_survey} shows the result of this analysis.
All surveys except for SDSS (optical surveys) independently recover the same U-shaped evolution: a shallow gradient at young ages, a steepening to ${\sim}-0.07$~dex\,kpc$^{-1}$ near $5$--$6$~Gyr, and a flattening to near zero for the oldest populations.
The agreement among these four is close throughout the intermediate-age range where they overlap, with the curves disagreeing at the oldest ages where the bins are sparse and the error bars large.
SDSS, on the other hand, contributes only $6{,}417$ stars and clears our fitting thresholds in just three age bins, all older than $8$~Gyr.
There is thus too little data to constrain the shape of the evolution across the full age range. 
Where there are measurements, SDSS traces the same flattening toward old ages as the other surveys, although it stays ${\sim}0.02$~dex\,kpc$^{-1}$ below the combined curve in the two oldest bins (${\sim}2\sigma$ offset), remaining slightly negative where the combined sample turns positive.

We again remind the reader that while this test rules out artifacts tied to differences between the surveys, it does not rule out biases shared by all surveys.

\section{Robustness to age errors}
\label{apdx:ages}

\begin{figure*}
    \color{blue}
    \centering
    \includegraphics[width=1\linewidth]{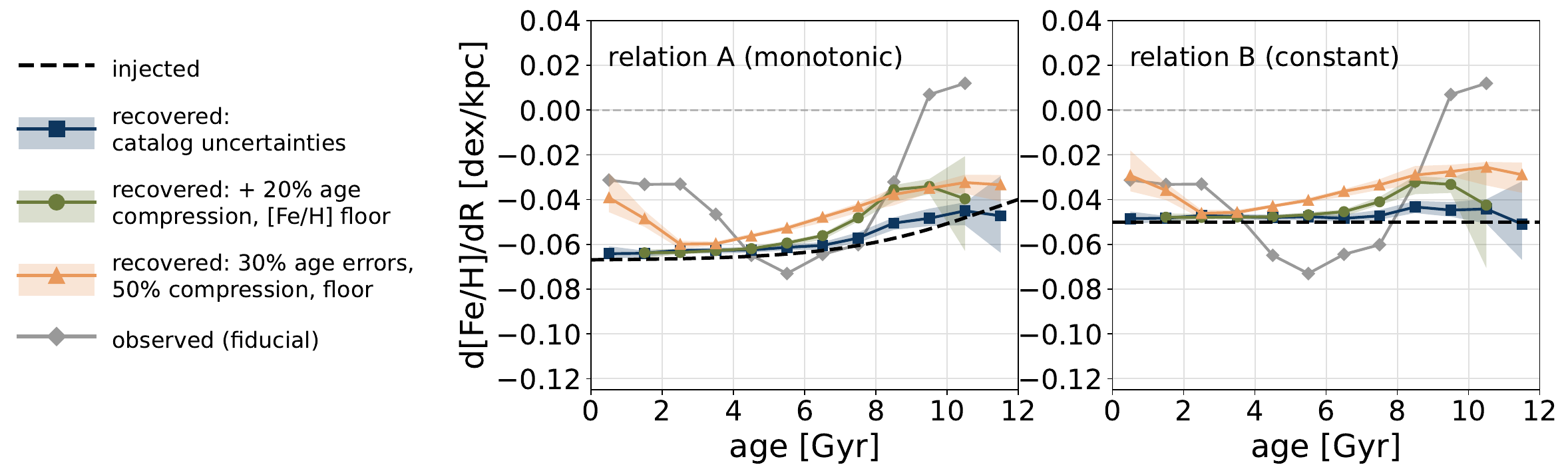}
    \caption{Tests of the age dependence of the radial [Fe/H] gradient against age errors. Injection tests for \textbf{(centre)} input relation A (the monotonic relation of \citealt{sun2026divergent}) and \textbf{(right)} input relation B (a constant $-0.05$~dex\,kpc$^{-1}$). The black dashed line is the injected relation, and the grey diamonds show the observed curve of Figure~\ref{fig:radial_gradient_vs_age} for comparison. The coloured curves are the median gradients recovered from 30 mock realizations, with bands showing the $16$--$84$th percentile range, for three error models: the catalog uncertainties (dark blue squares); the mild model, which adds a $20\%$ compression of the ages toward $5$~Gyr and removes stars with ${\rm [Fe/H]}<-0.8$ (green circles); and the pessimistic stress test, with $30\%$ age errors, an assumed $50\%$ compression toward $5$~Gyr, and the same floor (orange triangles). A point is shown only where at least half of the realizations yield a fit.}
    \label{fig:robustness_injection}
\end{figure*}

The Monte Carlo band of Figure~\ref{fig:radial_gradient_vs_age} tests only noise at the level of the catalog uncertainties, so here we ask whether larger or systematic age errors could turn a different underlying relation into the U-shape we measure (i.e., whether the U-shape is an artifact of age errors).
We keep the position and catalog age $\tau$ of every star as the truth and assign it a synthetic metallicity ${\rm [Fe/H]}=z_0(\tau)+g(\tau)\,(R-8~{\rm kpc})$ plus Gaussian noise, where $z_0(\tau)$ and the scatter ($0.06$--$0.28$~dex, increasing with age) are the median and standard deviation measured in the data at each age, and $g(\tau)$ is a known input relation: either (A) the monotonic thin-disk relation fitted by \citet{sun2026divergent}, or (B) a constant $-0.05$~dex\,kpc$^{-1}$.
We then perturb the distance, age, and [Fe/H] of each mock star and repeat the binning and fit of Figure~\ref{fig:radial_gradient_vs_age}, for 30 realizations (Figure~\ref{fig:robustness_injection}).
With the catalog uncertainties, both inputs are recovered to within $0.005$~dex\,kpc$^{-1}$ at every age.
The catalog ages are also compressed at the young and old ends (\citealt{Shen2026}) relative to open cluster ages by an unquantified amount, so we consider two assumed models.
Both models compress the ages toward $5$~Gyr, roughly where the catalog ages switch from overestimating to underestimating the open-cluster ages.
The mild model adds to the catalog uncertainties a pull of the observed ages $20\%$ of the way toward $5$~Gyr and removes mock stars with ${\rm [Fe/H]}<-0.8$, mimicking the metallicity floor of the catalog. The gradients at $8$--$11$~Gyr then flatten by $0.007$--$0.018$~dex\,kpc$^{-1}$, while those at $1$--$4$~Gyr change by less than $0.002$~dex\,kpc$^{-1}$.
The second model is a deliberately pessimistic stress test: we use Gaussian age errors of $30\%$, ages pulled $50\%$ of the way toward $5$~Gyr, and the same metallicity floor.
It flattens the gradients at $0$--$2$~Gyr by $0.014$--$0.025$~dex\,kpc$^{-1}$ and those at $8$--$11$~Gyr by $0.015$--$0.024$~dex\,kpc$^{-1}$, and turns both the monotonic and the constant input into a shallow U about $0.02$~dex\,kpc$^{-1}$ deep with its minimum at $2$--$4$~Gyr.
Errors of this kind could therefore account for a small part of the flattening we observe at either end, but not for its strength, for the strong U-shape with its minimum at $5$--$6$~Gyr, or for the near-zero gradient at old ages.

\section{Radial metallicity gradients as a function of age}
\label{apdx:gradient_table}

Table~\ref{tab:radial_gradients_age} lists the radial [Fe/H] gradients in $1$~Gyr age bins for the full sample and for the low-$\alpha$ and high-$\alpha$ sequences, which are plotted in Figures~\ref{fig:radial_gradient_vs_age} and \ref{fig:radial_gradient_thin_thick}.

\begin{table*}
  \centering
  \caption{Radial metallicity gradients as a function of age within $|Z|<1$~kpc, for the full sample and for the low-$\alpha$ and high-$\alpha$ sequences. The fits use 1~kpc radial bins over $6<R<14$~kpc, where we include bins with at least 50 stars, and $N$ is the number of stars in the fitted bins. Gradients are in dex\,kpc$^{-1}$, with the formal error of the linear fit to the bin medians.}
  \label{tab:radial_gradients_age}
  \small
  \setlength{\tabcolsep}{4pt}
  \begin{tabular}{ccccccc}
    \hline
     & \multicolumn{2}{c}{Full sample} & \multicolumn{2}{c}{Low-$\alpha$} & \multicolumn{2}{c}{High-$\alpha$} \\
    Age [Gyr] & $N$ & d[Fe/H]/d$R$ & $N$ & d[Fe/H]/d$R$ & $N$ & d[Fe/H]/d$R$ \\
    \hline
    0--1 & 2221 & $-0.0313 \pm 0.0096$ & 2221 & $-0.0313 \pm 0.0096$ & \nodata & \nodata \\
    1--2 & 15293 & $-0.0332 \pm 0.0044$ & 15292 & $-0.0332 \pm 0.0044$ & \nodata & \nodata \\
    2--3 & 29328 & $-0.0331 \pm 0.0029$ & 29300 & $-0.0332 \pm 0.0030$ & \nodata & \nodata \\
    3--4 & 41377 & $-0.0466 \pm 0.0030$ & 41317 & $-0.0466 \pm 0.0030$ & \nodata & \nodata \\
    4--5 & 51736 & $-0.0649 \pm 0.0047$ & 51097 & $-0.0660 \pm 0.0047$ & 225 & $-0.0090 \pm 0.0095$ \\
    5--6 & 56211 & $-0.0730 \pm 0.0043$ & 52432 & $-0.0785 \pm 0.0044$ & 2336 & $-0.0072 \pm 0.0012$ \\
    6--7 & 56484 & $-0.0644 \pm 0.0051$ & 45248 & $-0.0780 \pm 0.0037$ & 8891 & $-0.0156 \pm 0.0013$ \\
    7--8 & 51229 & $-0.0602 \pm 0.0049$ & 37044 & $-0.0761 \pm 0.0039$ & 11432 & $-0.0118 \pm 0.0032$ \\
    8--9 & 49297 & $-0.0321 \pm 0.0039$ & 20565 & $-0.0603 \pm 0.0052$ & 24079 & $-0.0142 \pm 0.0037$ \\
    9--10 & 48172 & $0.0069 \pm 0.0040$ & 5887 & $-0.0189 \pm 0.0099$ & 40280 & $0.0073 \pm 0.0036$ \\
    10--11 & 10599 & $0.0119 \pm 0.0062$ & 555 & $-0.0065 \pm 0.0082$ & 9801 & $0.0117 \pm 0.0049$ \\
    \hline
  \end{tabular}
\end{table*}

\section{Other $\alpha$ elements}
\label{apdx:alpha}

\begin{figure*}[th]
    \centering
    \includegraphics[width=1\linewidth]{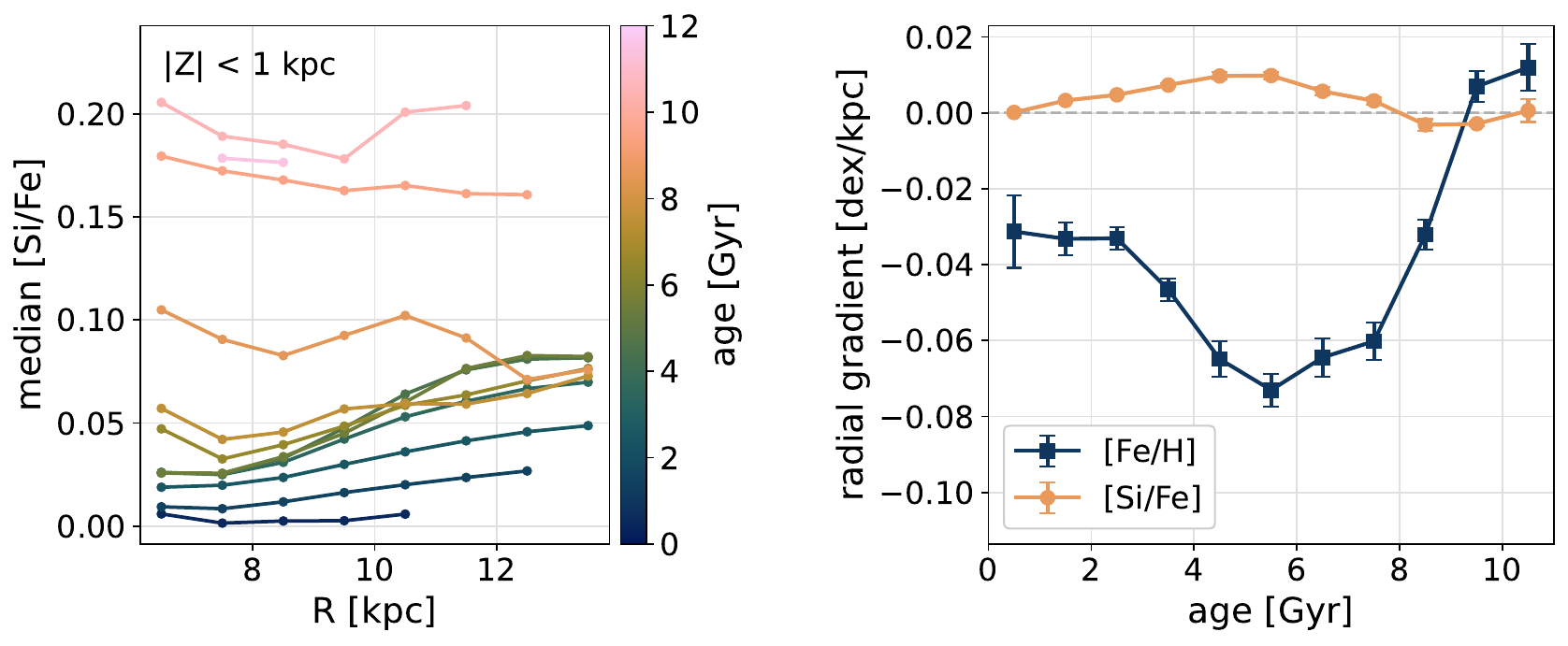}
    \caption{Same as the center and right panels of Figure~\ref{fig:radial_gradient_vs_age}, but for [Si/Fe].}
    \label{fig:sife_gradient}
\end{figure*}

\begin{figure*}[th]
    \centering
    \includegraphics[width=1\linewidth]{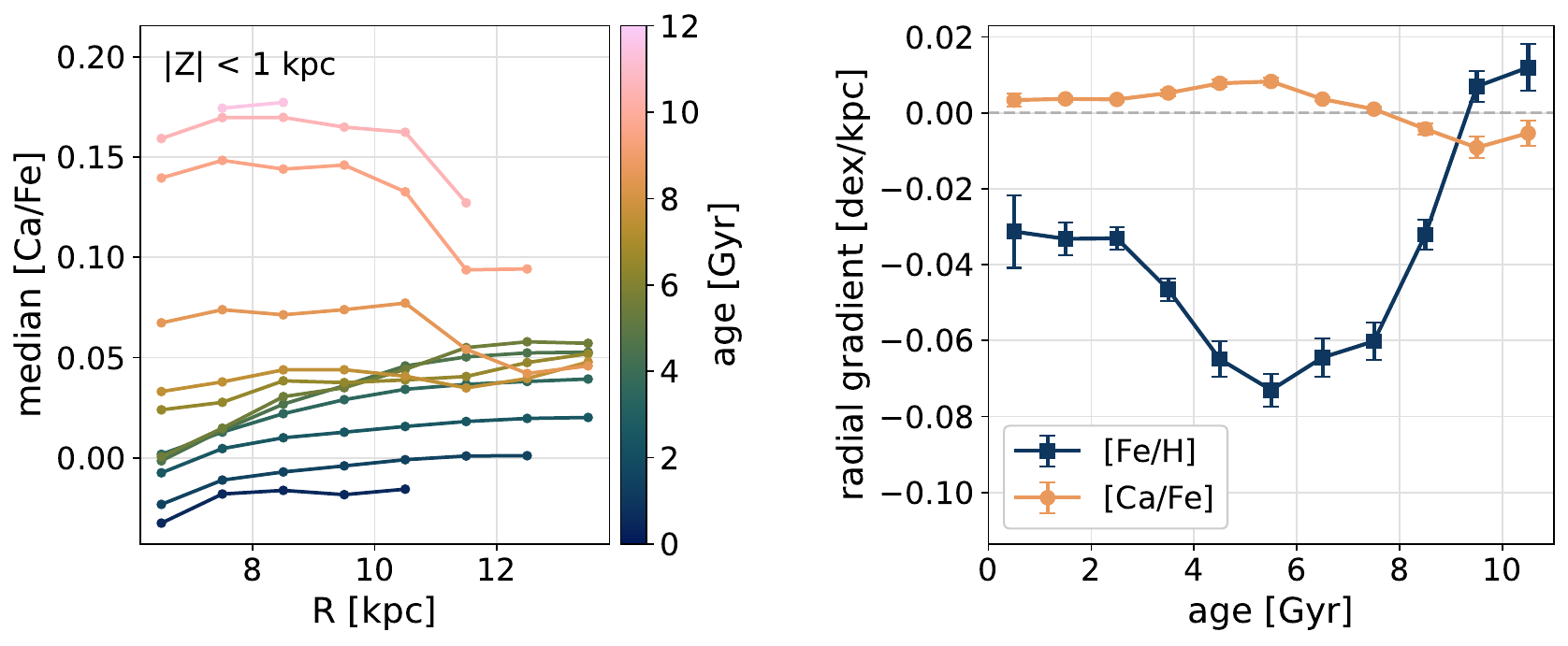}
    \caption{Same as the center and right panels of Figure~\ref{fig:radial_gradient_vs_age}, but for [Ca/Fe].}
    \label{fig:cafe_gradient}
\end{figure*}

\begin{figure*}[th]
    \centering
    \includegraphics[width=1\linewidth]{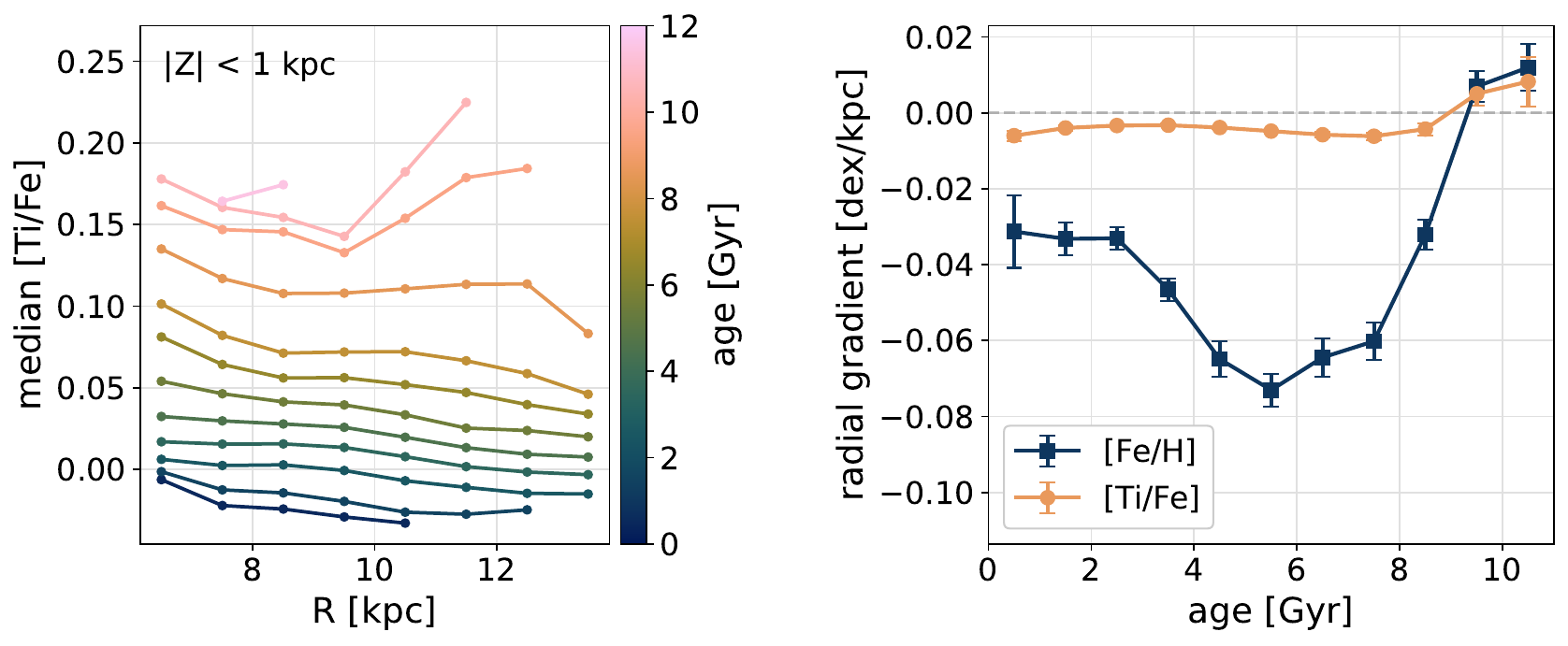}
    \caption{Same as the center and right panels of Figure~\ref{fig:radial_gradient_vs_age}, but for [Ti/Fe].}
    \label{fig:tife_gradient}
\end{figure*}

In the main text we used [Mg/Fe] as the fiducial $\alpha$ tracer (Figure~\ref{fig:radial_gradient_vs_age}). Figures~\ref{fig:sife_gradient}--\ref{fig:tife_gradient} show that the same behaviour holds for the other well-measured $\alpha$ elements: the radial gradients of [Si/Fe], [Ca/Fe], and [Ti/Fe] are all weak ($|{\rm d}[X/{\rm Fe}]/{\rm d}R|\lesssim0.01$~dex\,kpc$^{-1}$) and nearly age-independent, in contrast to the steep, U-shaped [Fe/H] gradient. The flatness of the [$\alpha$/Fe] gradient is therefore robust to the choice of $\alpha$ element.

\bibliography{references, references2}{}
\bibliographystyle{aasjournalv7}

\end{CJK}

\end{document}